\documentclass[graybox]{svmult}

\usepackage{type1cm}        
\usepackage{makeidx}         
\usepackage{graphicx}        
\usepackage{multicol}        
\usepackage[bottom]{footmisc}

\usepackage{newtxtext}       %
\usepackage{newtxmath}       

\usepackage[latin1]{inputenc}
\usepackage{graphicx}
\usepackage{epsfig}
\usepackage[normalsize,it,hang]{subfigure}
\DeclareGraphicsExtensions{.jpg,.png,.jpg,.jpg}

\makeindex             

\begin{document}

\title*{
{\small \tt $3$rd DECOD Workshop \\ (DElays and COnstraints in Distributed parameter systems)  \\  23-26 November 2021, Gif-sur-Yvette, France}\vspace{2cm} \\
Ramp metering: \\ modeling, simulations and control issues}
\author{C\'{e}dric Join, Hassane Aboua\"{\i}ssa and Michel Fliess}
\institute{C\'{e}dric Join \at CRAN (CNRS, UMR 7039), Universit\'{e} de Lorraine, BP 239, 54506 Vand{\oe}uvre-l\`{e}s-Nancy, France.\\\email{cedric.join@univ-lorraine.fr}
\and Hassane Aboua\"{\i}ssa  \at Univ. Artois, UR 3926, Laboratoire de G\'enie Informatique et d'Automatique de l'Artois (LGI2A) F-62400  B\'{e}thune, France.\\ \email{hassane.abouaissa@univ-artois.fr}
\and Michel Fliess \at LIX (CNRS, UMR 7161), \'Ecole polytechnique, 91128 Palaiseau, France.\\ \email{Michel.Fliess@polytechnique.edu}
\and C\'{e}dric Join and Michel Fliess \at AL.I.E.N., 7 rue Maurice Barr\`{e}s, 54330 V\'{e}zelise, France.\\\email{ \{cedric.join, michel.fliess\}@alien-sas.com}
}


%
%
\maketitle

\abstract*{Each chapter should be preceded by an abstract (no more than 200 words) that summarizes the content. The abstract will appear \textit{online} at \url{www.SpringerLink.com} and be available with unrestricted access. This allows unregistered users to read the abstract as a teaser for the complete chapter.
Please use the 'starred' version of the \texttt{abstract} command for typesetting the text of the online abstracts (cf. source file of this chapter template \texttt{abstract}) and include them with the source files of your manuscript. Use the plain \texttt{abstract} command if the abstract is also to appear in the printed version of the book.}

\abstract{The aim of ramp metering is to improve the highway traffic conditions by an appropriate regulation of the inflow from the on-ramps to the highway mainstream. Our presentation rests on several improvements: 1) Our simulation techniques do not need contrarily to other approaches any heuristic fundamental law. 2) There is no need of crucial time-varying quantities, like the critical density, which is most difficult to estimate correctly online. 3) Our feedback loop, which is stemming from model-free control, is easy to implement and shows an excellent robustness with respect to model mismatch. Several computer experiments are displayed and discussed.}


\section{Introduction}
\label{intro}
\emph{Ramp metering} is about the use of traffic signals at highway on-ramps in order to control the rate of vehicles entering the highway. The signals can be set for different metering rates to optimize freeway flow and minimize congestion. The social and ecological damages due to traffic jams justify the huge academic literature which has been devoted to traffic flow dynamics and its regulation. See, \textit{e.g.}, \cite{kachroo1,kachroo2,kerner1,kerner2,mammar,may,papageorgiou0,treiber}, and references therein. Lack of space prevents us from a careful analysis of the various standpoints which are often antagonist.

Our contributions may be summarized like follows:
\begin{enumerate}
\item {\bf No fundamental diagram for simulation purposes.} Before being implemented in practice any ramp metering procedure ought to be tested via computers. The similarities between highway traffic and hydrodynamics explain the importance of the  Lighthill-Whitham-Richards (LWR) first order partial differential equation \cite{LWR1,LWR2}. Its numerical integration necessitates to know the relationship between the traffic density and the traffic speed. Several empirical laws, called \emph{fundamental diagrams}, have been proposed. The calibration of those diagrams is far from being obvious. That is why contrarily to other approaches (see, \textit{e.g.}, \cite{barcelo}) we ignore them. Fundamental diagrams are replaced by traffic data which are collected during various situations. Note moreover that our viewpoint, contrarily to other ones like METANET \cite{kots}, does not necessitate higher order partial differential equations.
\item {\bf No difficult estimation technique.} Implementing any control law requires crucial quantities like the \emph{critical density}, which are time-varying and therefore almost impossible to estimate in real time. They are replaced by {\it ad hoc} quantities which are deduced at once from easily measurable data.
\item {\bf Model-free control.} ALINEA\footnote{It is the acronym of \textit{\underline{A}sservissement \underline{LIN}\'{e}aire d'\underline{E}ntr\'{e}e \underline{A}utorouti\`{e}re}.} \cite{hist1,hist2,papageorgiou1,papageorgiou2} is perhaps the most popular feedback control algorithm for traffic regulation. After first trials in Paris it has been quite often employed in many different places. Following \cite{sofia}, we are using here \emph{model-free control} in the sense of \cite{mfc13,nicu} and, more precisely, \emph{intelligent proportional controllers}. This setting, which has been successfully tested in many concrete situations, has already been illustrated via various questions about intelligent transportation systems (see, \textit{e.g.}, \cite{andrea1,baciu,haddar,menhour,milan,polack,villagra,wang,yang}).  Concrete experiments show today that model-free control not only yields better traffic regulation than ALINEA but is also simpler to implement.\footnote{There are many variants of ALINEA in the literature. We have selected for our computer comparisons the version \cite{alinea0}.}
\end{enumerate}
Some remarks might be useful:
\begin{itemize}
\item Let us emphasize that the LWR partial differential equation is only related to simulation purposes and not to traffic control. Such a control would necessitate a real-time calibration, which seems today beyond all reasonable hope.
\item Achieving an efficient traffic regulation is simpler than obtaining reliable computer simulations. This ascertainment  remains valid in  most applications where model-free control plays a key r\^{o}le. 
\end{itemize}
Our paper is organized as follows. Section \ref{mfc} reviews model-free control and the corresponding \emph{intelligent proportional controller}. Following \cite{alinea}, Section \ref{prop} investigates the relationship with proportional controllers, which should be considered as the backbone of ALINEA. Modeling issues are discussed in Section \ref{rm}: The partial differential equation stemming from elementary hydrodynamic conservation laws, some fundamental empirical laws, the space discretization and its use without any fundamental empirical law, the ALINEA and model-free regulation without the critical density. Computer simulations are displayed and commented in Section \ref{simu}. Section \ref{sudd} in particular exhibits the superiority of model-free control in the presence of sudden changes. Section \ref{conclu} contains some concluding remarks.

\section{Model-free control}\label{mfc}
\subsection{A short review}
\subsubsection{Ultra-local model}\label{ulm}
Consider only, for simplicity's sake, SISO (single-input single-output) systems.
Elementary functional analysis and differential algebra as used in \cite{mfc13} show as well as practical experiments  that most, or at least many, concrete systems may be approximated by the \emph{ultra-local} 
\begin{equation}
\boxed{\dot{y} = F + \alpha u}
\label{1}
\end{equation}
where
\begin{itemize}
\item the control and output variables are respectively $u$ and $y$,
\item the constant $\alpha \in \mathbb{R}$ is chosen by the practitioner such that $\alpha u$ and
$\dot{y}$ are of the same magnitude. Therefor $\alpha$ does not need to be precisely estimated.
\end{itemize}
The following comments might be useful:
\begin{itemize}
\item $F$ is estimated via the knowledge of the control and output variables $u$ and $y$,
\item $F$ subsumes not only the unknown structure of the system, but also
any external disturbance.
\end{itemize}

\subsubsection{Intelligent controllers}
Close the loop with the following \emph{intelligent proportional controller}, or \emph{iP},
\begin{equation}\label{ip}
\boxed{u = - \frac{F - \dot{y}^\ast + K_P e}{\alpha}}
\end{equation}
where:
\begin{itemize}
\item $y^\ast$ is the reference trajectory,
\item $e = y - y^\ast$ is the tracking error,
\item $K_P$ is a tuning gain.
\end{itemize}
Combining Equations \eqref{1} and \eqref{ip} yields:
\begin{equation*}\label{dyna}
\dot{e} + K_P e = 0
\end{equation*}
where $F$ does not appear anymore. Thus $\lim_{t \to +\infty} e(t) = 0$ iff $K_P > 0$. This local stability property proves that the tuning of $K_P$ is straightforward. This is a major difference with the classic gain tuning for PIs and PIDs (see, \textit{e.g.}, \cite{astrom} and the references therein).

\subsubsection{Estimation of $F$}\label{F}
Under a weak integrability condition, $F$ in Equation \eqref{1} may be ``well'' approximated by a piecewise constant function $F_{\text{est}}$ (see, \textit{e.g.}, \cite{bourbaki}). The estimation techniques below are borrowed 
from \cite{sira1,sira2,sira}.\footnote{They were often used in practice for parameter identification (see, \textit{e.g.}, \cite{train}).} Let us summarize two types of computations:
\begin{enumerate}
\item Rewrite Equation \eqref{1}  in the operational domain (see, \emph{e.g.}, \cite{yosida}): 
$$
sY = \frac{\Phi}{s}+\alpha U +y(0)
$$
where $\Phi$ is a constant. We get rid of the initial condition $y(0)$ by deriving both sides with respect to $s$:
$$
Y + s\frac{dY}{ds}=-\frac{\Phi}{s^2}+\alpha \frac{dU}{ds}
$$
Noise attenuation is achieved by multiplying both sides on the left by $s^{-2}$, since integration with respect to time is a lowpass filter (see \cite{noise} for further details). It yields in the time domain the realtime estimate, 
thanks to the equivalence between $\frac{d}{ds}$ and the multiplication by $-t$,
\begin{equation}\label{integral}
{\small F_{\text{est}}(t)  =-\frac{6}{\tau^3}\int_{t-\tau}^t \left\lbrack (\tau -2\sigma)y(\sigma)+\alpha\sigma(\tau -\sigma)u(\sigma) \right\rbrack d\sigma }
\end{equation}
where $\tau > 0$ might be quite small. This integral may of course be replaced in practice by a classic digital filter.
\item Close the loop with the iP \eqref{ip}. It yields:
$$
F_{\text{est}}(t) = \frac{1}{\tau}\left[\int_{t - \tau}^{t}\left(\dot{y}^{\ast}-\alpha u
- K_P e \right) d\sigma \right] 
$$
\end{enumerate}

\subsection{PI and iP}\label{PiP}
Consider the classic proportional-integral controller, or PI,
\begin{equation}\label{cpi}
 \boxed{u (t) = k_p e(t) + k_i \int e(\tau) d\tau}
\end{equation}
where $k_p, k_i \in \mathbb{R}$ are constants. A crude sampling of the integral $\int e(\tau) d\tau$ through a
Riemann sum ${\cal{I}}(t)$ leads to
$$
\int e(\tau) d\tau \simeq  {\cal{I}}(t) = {\cal{I}}(t-h) + h e(t)
$$
where $h$ is the sampling interval. The corresponding discrete form
of Equation \eqref{cpi} reads:
$$
u(t) = k_p e(t) + k_i {\cal{I}}(t) = k_p e(t) + k_i {\cal{I}}(t-h) + k_i h e(t)
$$
Combining the above equation with $$u(t-h) = k_p e(t-h) + k_i
{\cal{I}}(t-h)$$ yields
\begin{equation}
\label{eqPIRiemannDiscrSix} u(t) = u(t - h) + k_p \left( e(t) - e(t
- h) \right) + k_i h e(t)
\end{equation}
Replace in Equation \eqref{ip} $F$ by ${\dot y}(t) - \alpha u (t-h)$ and therefore by
\begin{equation}
\frac{y(t) - y(t-h)}{h} - \alpha u (t-h)
\label{Fest}
\end{equation}
It yields
\begin{equation}
\label{eqDiscr_i-POne} u (t) = u (t - h) - \frac{e(t) -
e(t-h)}{h\alpha} - \dfrac{K_P}{\alpha}\, e(t)
\end{equation}
Equations \eqref{eqPIRiemannDiscrSix} and
\eqref{eqDiscr_i-POne} become {\bf identical} if we set
\begin{align}
\label{eqPI_i-P_corresp} k_p &= - \dfrac{1}{\alpha h}, \quad k_i =
-\dfrac{K_P}{\alpha h}
\end{align}
Let us emphasize that this important result, which was first stated in \cite{andrea,mfc13}, is only valid in discrete time: Formulae \eqref{eqPI_i-P_corresp} become meaningless if $h \downarrow 0$.

\subsection{Application to integral controllers}\label{prop}
Set in Equation \eqref{cpi} $k_p=0$. It yields the integral controller, or I,
\begin{equation}\label{ci}
 \boxed{u (t) = k_i \int e(\tau) d\tau}
\end{equation}
Derive both sides in Equation \eqref{cpi}:
$$
\dot{u} = k_p \dot{e} + k_i e
$$
It shows that the PI and I controllers are ``close'' if $\dot{e}$ remains ``small.'' It implies of course that the initial condition $y(0)$ is ``close'' to the initial point $y^\ast (0)$ of the reference trajectory. Let us try to explain why:
\begin{itemize}
\item the feedback loop is supposed to render $y - y^\ast$ quickly as small as possible,
\item the derivative $\dot{e}$ might therefore be large. 
\end{itemize}
 It follows from Equations \eqref{eqPIRiemannDiscrSix} and \eqref{Fest} that the sampled versions of the iP \eqref{ip} and the I \eqref{ci} are ``close'' (see \cite{alinea} for more details) if 
\begin{itemize}
\item the reference trajectory $y^\ast$ is ``slowly'' varying, and starts at the initial condition $y(0)$ or, at least, at a point which is quite close to it,
\item the disturbances and the corrupting noises are rather mild.
\end{itemize}

\section{Application to ramp metering}\label{rm}
\subsection{Traffic flow modeling}
\subsubsection{The LWR partial differential equation}\label{pde}
The Lighthill-Whitham-Richards, or LWR, partial differential equation \cite{LWR1,LWR2}, which was derived 65 years ago, is mimicking hydrodynamic conservation laws. It is the simplest model for a macroscopic traffic flow model
\begin{equation}\label{lwr}
\frac{\partial}{\partial t}\rho(t,x) +  \frac{\partial }{\partial x}f(t,x) = 0
\end{equation}
\begin{itemize}
\item $\rho$ is the traffic density,
\item the traffic flux $f = \rho \times v$ is the product of the traffic density and of the (mean) traffic speed $v$.
\end{itemize}
\subsubsection{Fundamental laws}\label{law}
There are many empirical laws relating $\rho$ and $v$. The Greenshield model \cite{green}
\begin{equation}\label{greenshield}
v(\rho) = v_f \left(1 - \frac{\rho}{\rho_m}\right)
\end{equation}
and the May formula \cite{may}:
\begin{equation}\label{mayeq}
v(\rho) = v_f \exp \left(- \frac{1}{a} \left(\frac{\rho}{\rho_c}\right)^a\right)
\end{equation}
for instance are quite popular. We will not try here to define the corresponding parameters.  Note however that they depend on $t$ and $x$. Their estimation is therefore rather cumbersome.

\subsubsection{A space discretization for numerical simulations}\label{discret}
\begin{figure*}[!ht]
\centering%
{\epsfig{figure=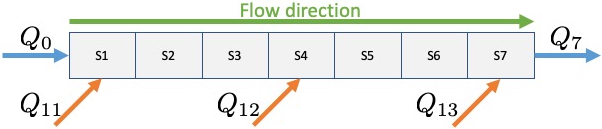,width=.8\textwidth}}
\caption{Segments and ramps of motorway portion}\label{R7}
\end{figure*}

Figure \ref{R7} displays a portion of a motorway in France, with $7$ segments $S\iota$, $\iota = 1, \dots, 7$. There are $3$ lanes on  $S\iota$, $\iota = 1, \dots, 6$, and $4$ on $S_7$. The lengths of the various segments are respectively $4.7, 0.6, 1.4, 1.7, 3.7, 0.6, 0.9$ ([km]). Write $Q_0$ (resp. $Q_7$) the inlet (resp. output) flow. The traffic flow on $S\iota$ is defined by $(Q_\iota, To_\iota, V_\iota)$, where $Q_\iota$, $T{o_\iota}$ and $V_\iota$ are respectively the flow rate, the occupation rate and the speed. Our aim is to regulate the flow rates $Q_{11}$, $Q_{12}$, $Q_{13}$ on the $3$ access ramps in order to ease the traffic flow on the motorway. The control variables are the green light durations $GD_\kappa$, $\kappa = 1, 2, 3$, on the $3$ on-ramps. The duration of a cycle of the ramp signals is $40$s. Set $15{\rm s} \leq GD_\kappa \leq 29{\rm s}$.


The vehicle conservation principle for each segment $\iota$ should be understood as a space discretization of Equation \eqref{lwr}. It reads
$$
\dot\rho_\iota = \frac{1}{L_\iota}\left(Q_{\iota-1}-Q_{\iota}\right)
$$
\begin{itemize}
\item $\rho_\iota$ is the density ([Veh/km]), 
\item $L_\iota$ is the length ([km]), 
\item $Q_{\iota-1}$ is the inlet flow ([Veh/min]): it is the sum of the upstream flow and of the ramp flow if any, 
\item $Q_{\iota}$ is the output flow ([Veh/min]).
\end{itemize}
Define the \emph{occupancy} rate ([\%]) by 
$$To_\iota = \frac{\rho_\iota}{\rho_{\iota, {\rm max}}} \times 100$$  
where $\rho_{\iota, {\rm max}}=\frac{\lambda_\iota}{{\tt size}}$, $\lambda_\iota$ is the number of lanes, ${\tt size}$ is the mean length ([km]) of a vehicle (here $5.5.10^{-3}$).

\begin{figure*}[!h]
\centering
\subfigure[$v_1=\mathcal{D}_1(To_1)$]{\rotatebox{-0}{\epsfig{figure=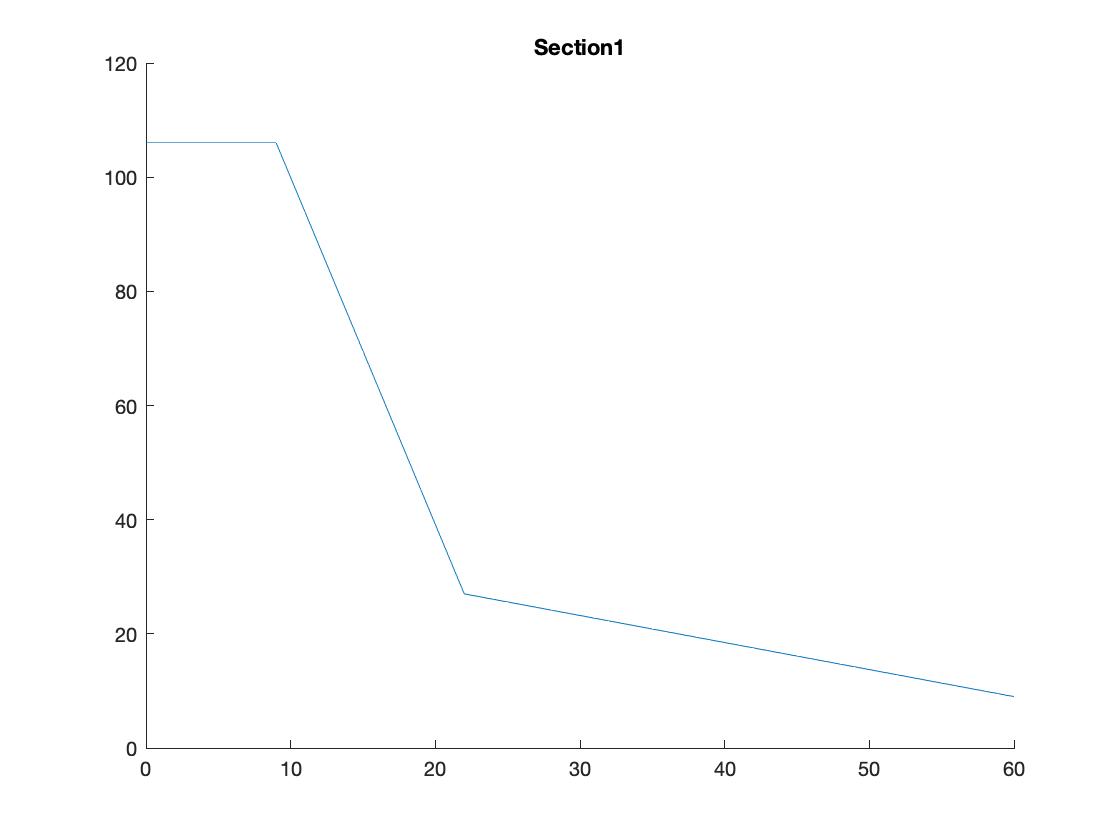,width=0.3\textwidth}}}
\subfigure[$v_2=\mathcal{D}_2(To_2)$]{\rotatebox{-0}{\epsfig{figure=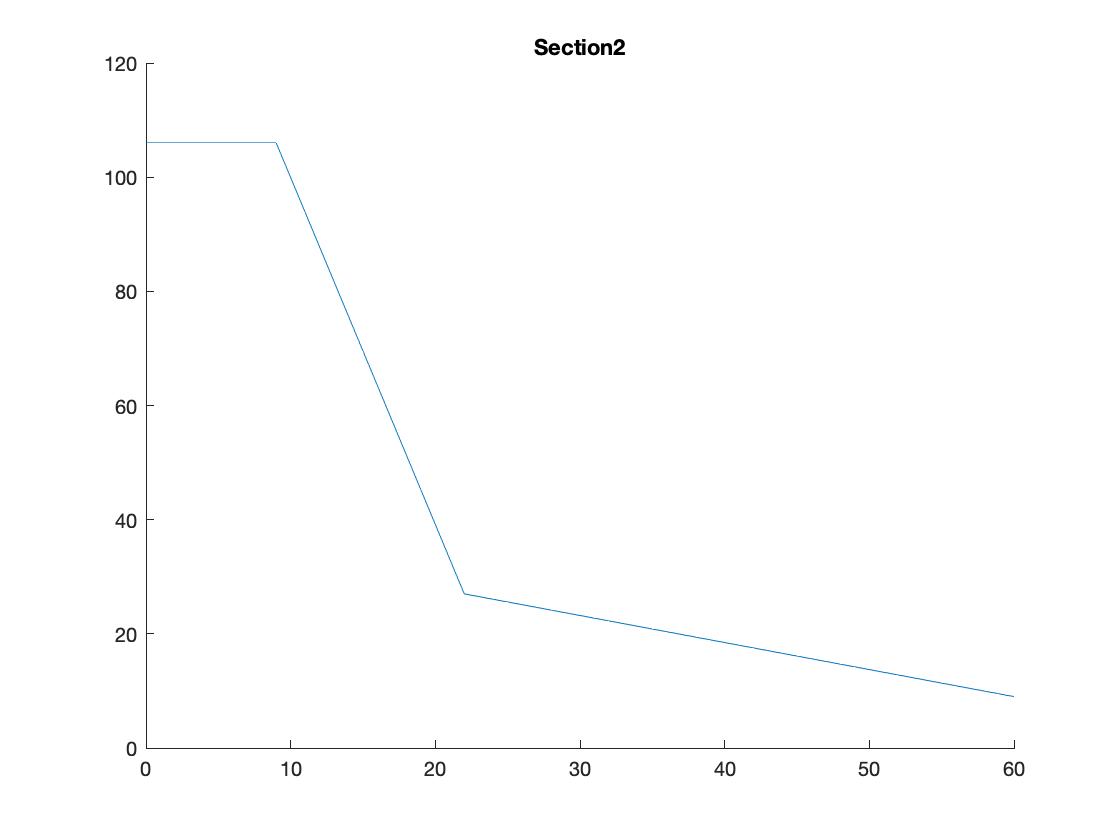,width=0.3\textwidth}}}
\subfigure[$v_3=\mathcal{D}_3(To_3)$]{\rotatebox{-0}{\epsfig{figure=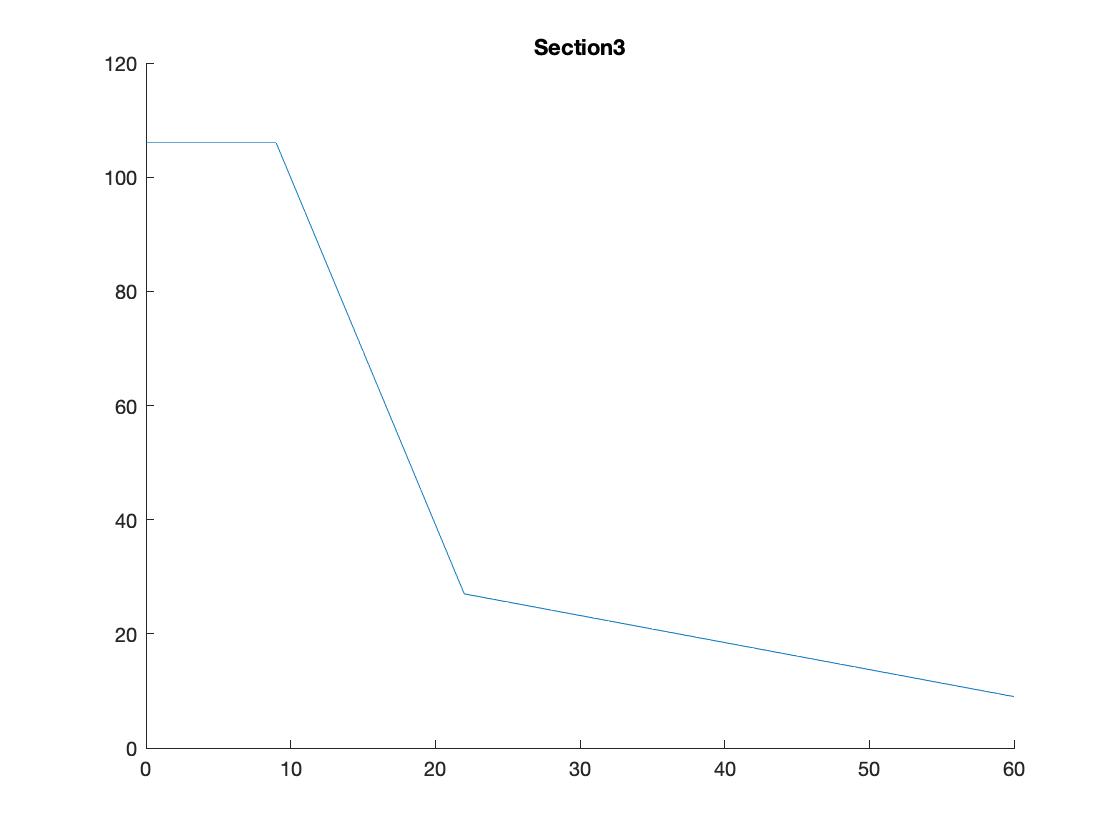,width=0.3\textwidth}}}
\subfigure[$v_4=\mathcal{D}_4(To_4)$]{\rotatebox{-0}{\epsfig{figure=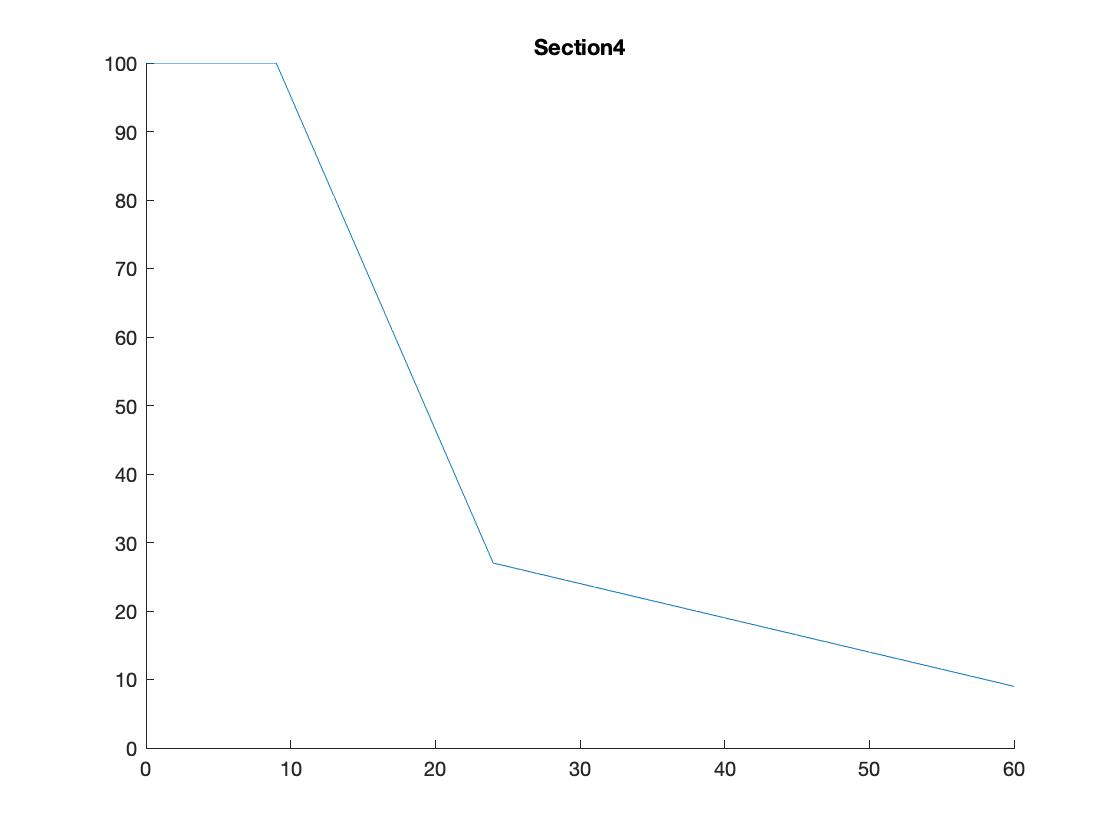,width=0.3\textwidth}}}
\subfigure[$v_5=\mathcal{D}_5(To_5)$]{\rotatebox{-0}{\epsfig{figure=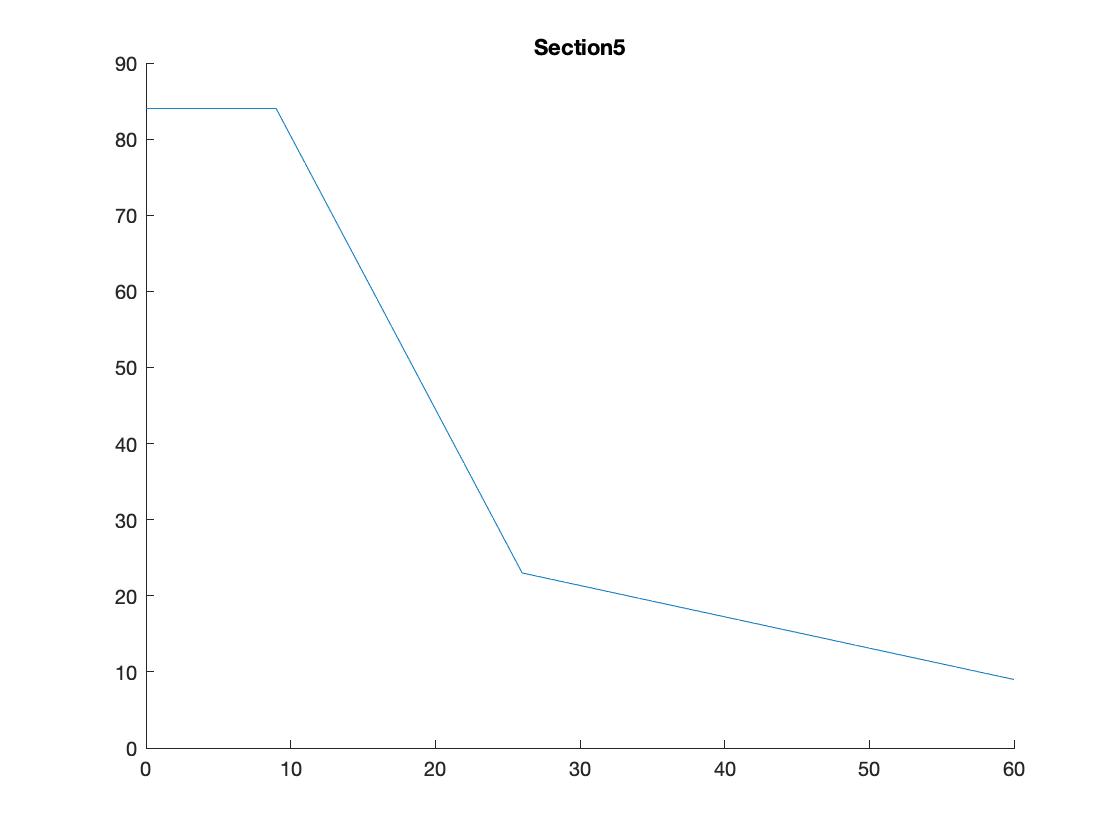,width=0.3\textwidth}}}
\subfigure[$v_6=\mathcal{D}_6(To_6)$]{\rotatebox{-0}{\epsfig{figure=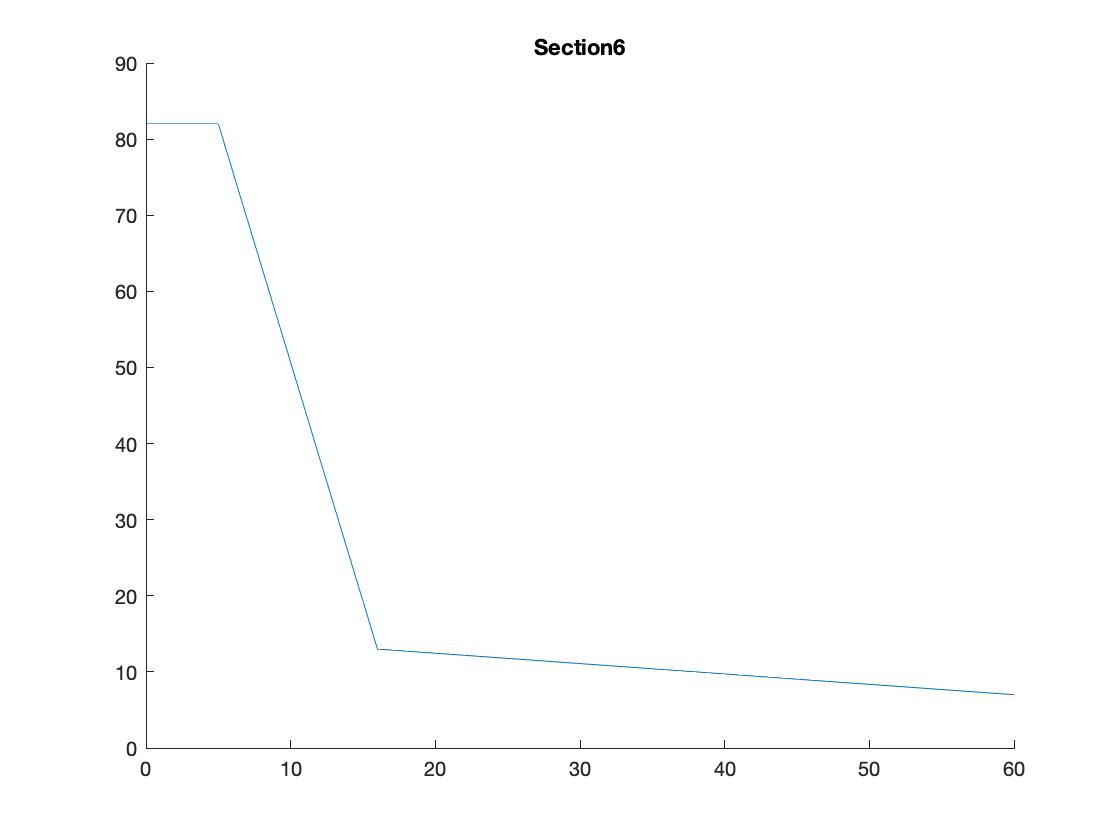,width=0.3\textwidth}}}
\subfigure[$v_7=\mathcal{D}_7(To_7)$]{\rotatebox{-0}{\epsfig{figure=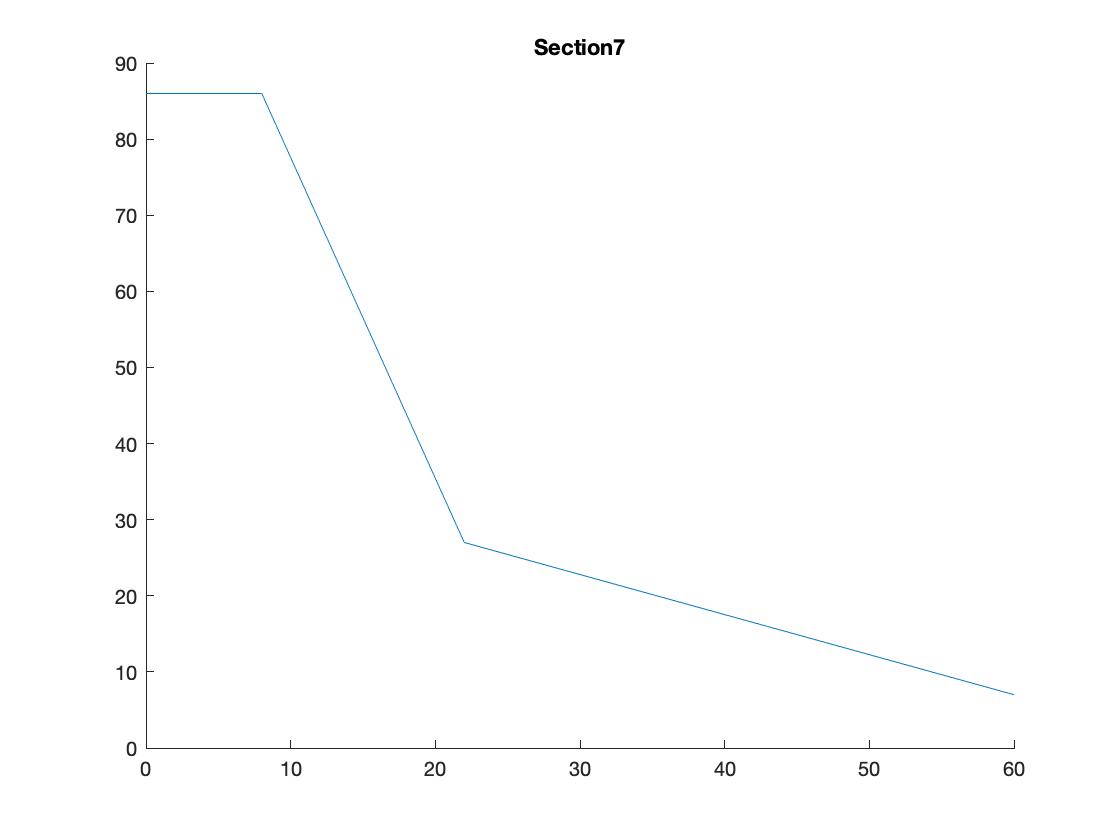,width=0.3\textwidth}}}
\caption{Diagrams}\label{Di}
\end{figure*}

The fundamental diagrams in Figure \ref{Di} relate the (mean) speed ([km/h]) and the occupancy, \textit{i.e.}, $v_i=\mathcal{D}_\iota(To_\iota)$. They are derived from real data which are, therefore, confidential..

\subsection{Feedback loops}

\subsubsection{ALINEA  without calibration}
Equation \eqref{ci} yields ALINEA in ramp metering. For $\kappa = 2$ it reads\footnote{The cases $\kappa = 1$ and $\kappa = 3$ are similar.}
$$GD_2=-K_{I,2} \int_0^t(To_4-To^\ast_4)d\tau$$
where the gain $K_{I,2} \in \mathbb{R}$ is set equal to $1$. A classic {\em anti-windup} setting (see, \textit{e.g.}, \cite{astrom}) is mandatory. The \emph{critical} occupancy rate $To^\ast_4$ corresponds to the \emph{critical} density $\rho_c$ in Formula \eqref{mayeq}. A ``good'' real-time calibration of $To^\ast_4$  seems today out of reach. We therefore set as in \cite{alinea0}
$$
\begin{cases}
To^\ast_4(t+1)=To^\ast_4(t)+\delta^+\text{ if }V_4 >V_{4, \text{threshold}}\\
To^\ast_4(t+1)=To^\ast_4(t)-\delta^- \text{ if not}
\end{cases}
$$
where
\begin{itemize}
\item $V_4$ is the mean speed on the 4th segment,
\item $V_{4, \text{threshold}} = V_{4, f} - 10$,
\item $V_{4, f}$ is the \emph{free} speed, \textit{i.e.}, the maximum speed when the traffic is light,
\item $\delta^+=0.15$, $\delta^-=0.3$.
\end{itemize}

\subsubsection{iP}\label{iptraffic}
Equation \eqref{1} reads here 
\begin{align*}
\dot To_1 &= F_1 + \alpha_1 GD_1 \\
\dot To_4 &= F_2 + \alpha_2 GD_2 \\
\dot To_7 &= F_3 + \alpha_3 GD_3
\end{align*}
The analogous feedback loops of Equation \eqref{ip} become
\begin{align*}
GD_1 &= \frac{\dot{To}_1^\ast - F_1 - \mathcal{K}_1(To_1 - To^\ast_1)}{\alpha_1} \\
GD_2 &= \frac{\dot{To}_4^\ast - F_2 - \mathcal{K}_2 (To_4 - To^\ast_4)}{\alpha_2} \\
GD_3 &= \frac{\dot{To}_7^\ast - F_3 - \mathcal{K}_3 (To_7 - To^\ast_7)}{\alpha_3}
\end{align*}
Set  $\alpha_j  = 30$, $K_{p,j} = 0.5$. $To^\ast_\kappa$, $\kappa = 1, 2, 3$ is replaced as above.

\section{Simulations}\label{simu}
\subsection{Generalities}
\begin{figure*}[!h]
\centering
\subfigure[Flow in \textit{[Veh/min]}]{\rotatebox{-0}{\epsfig{figure=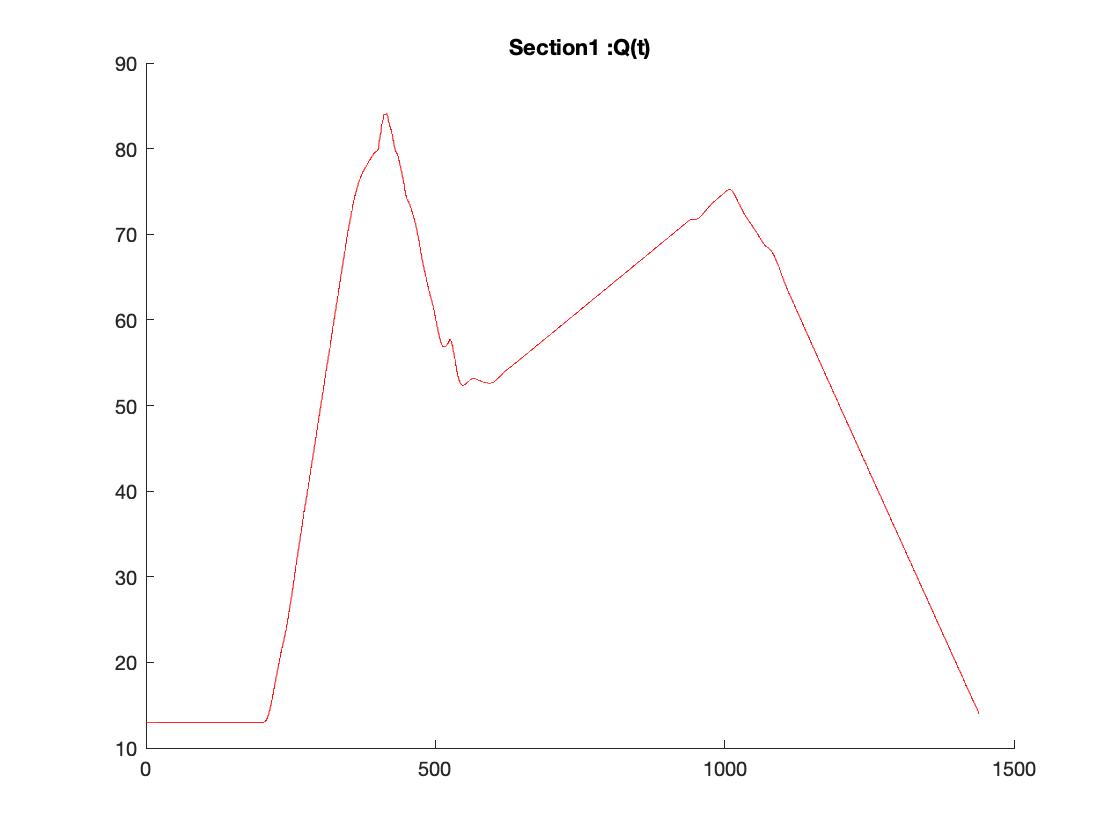,width=0.3\textwidth}}}
\subfigure[Occupancy rate in \textit{\%}]{\rotatebox{-0}{\epsfig{figure=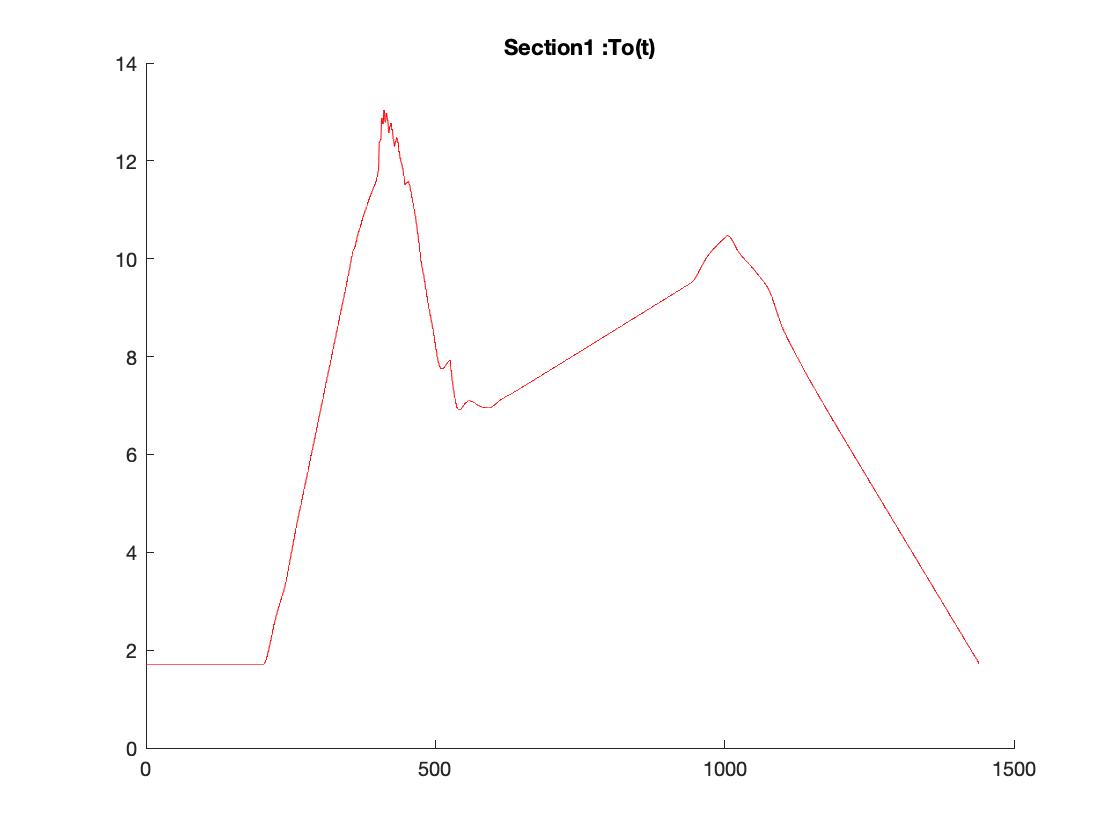,width=0.3\textwidth}}}
\subfigure[Speed in \textit{[km/h]}]{\rotatebox{-0}{\epsfig{figure=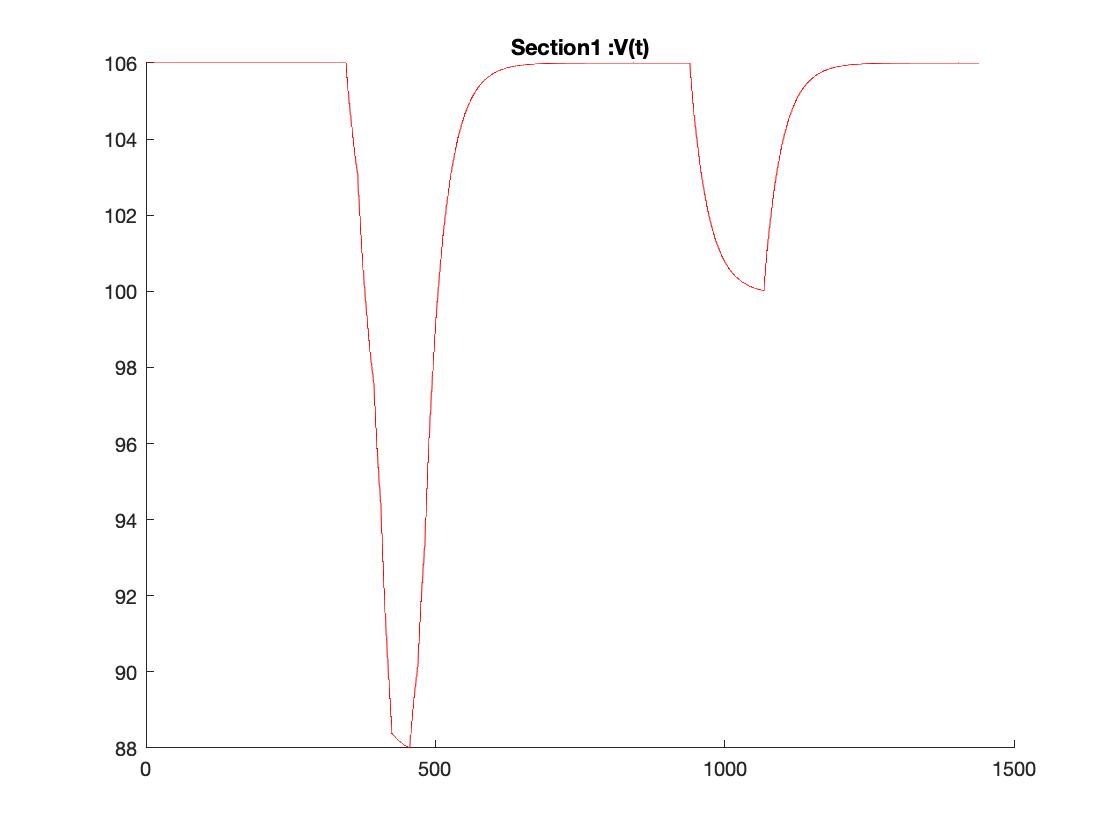,width=0.3\textwidth}}}
\caption{Section 1}\label{CSM1}
\end{figure*}

\begin{figure*}[!h]
\centering
\subfigure[Flow in \textit{[Veh/min]}]{\rotatebox{-0}{\epsfig{figure=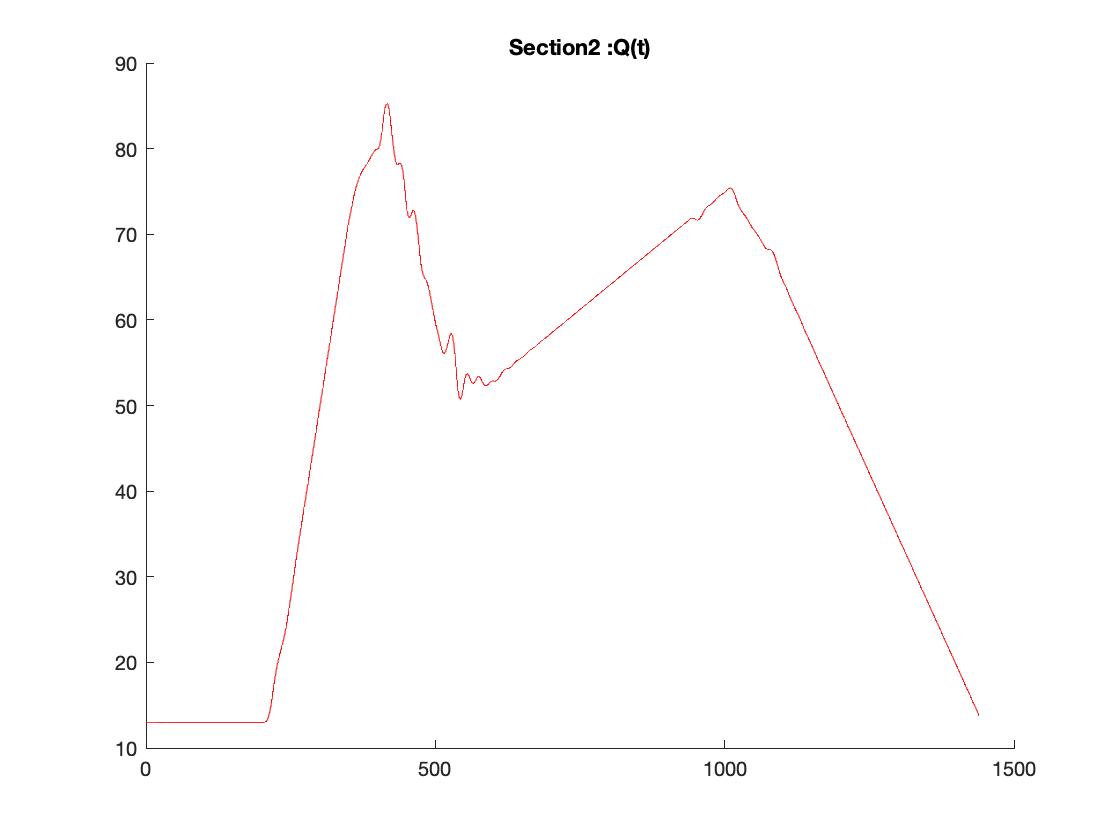,width=0.3\textwidth}}}
\subfigure[Occupancy rate in \textit{\%}]{\rotatebox{-0}{\epsfig{figure=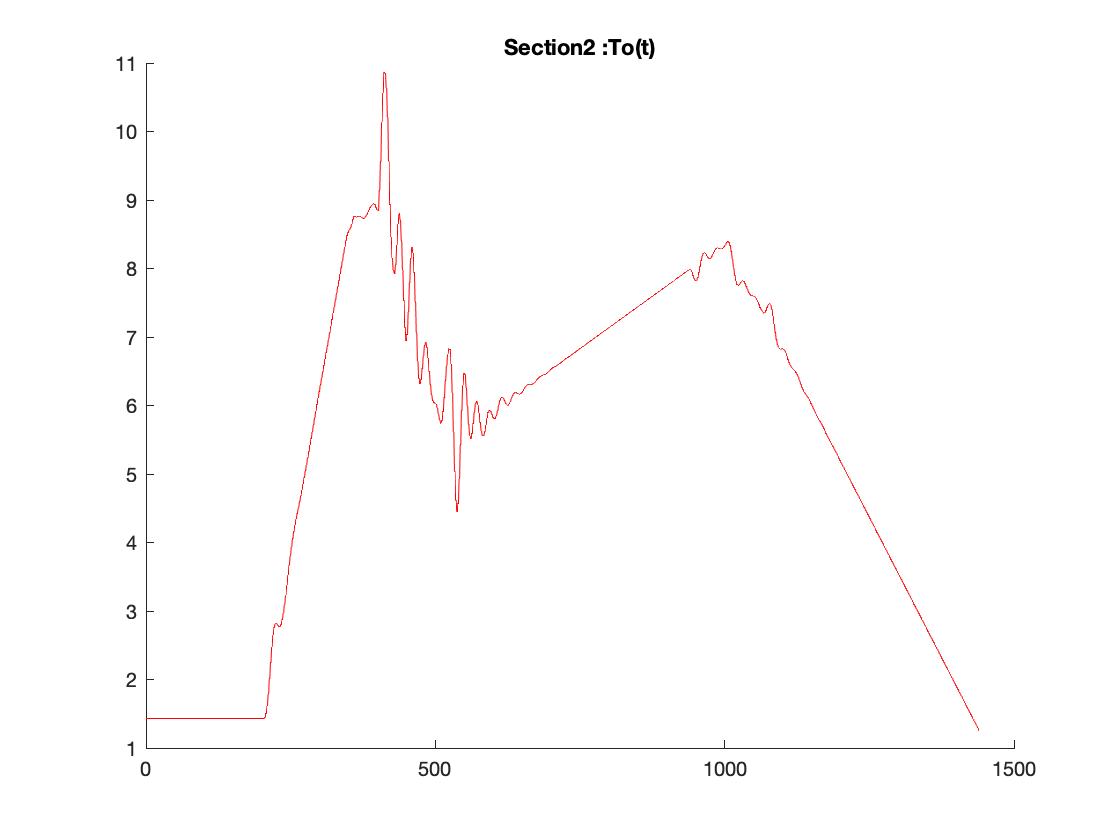,width=0.3\textwidth}}}
\subfigure[Speed in \textit{[km/h]}]{\rotatebox{-0}{\epsfig{figure=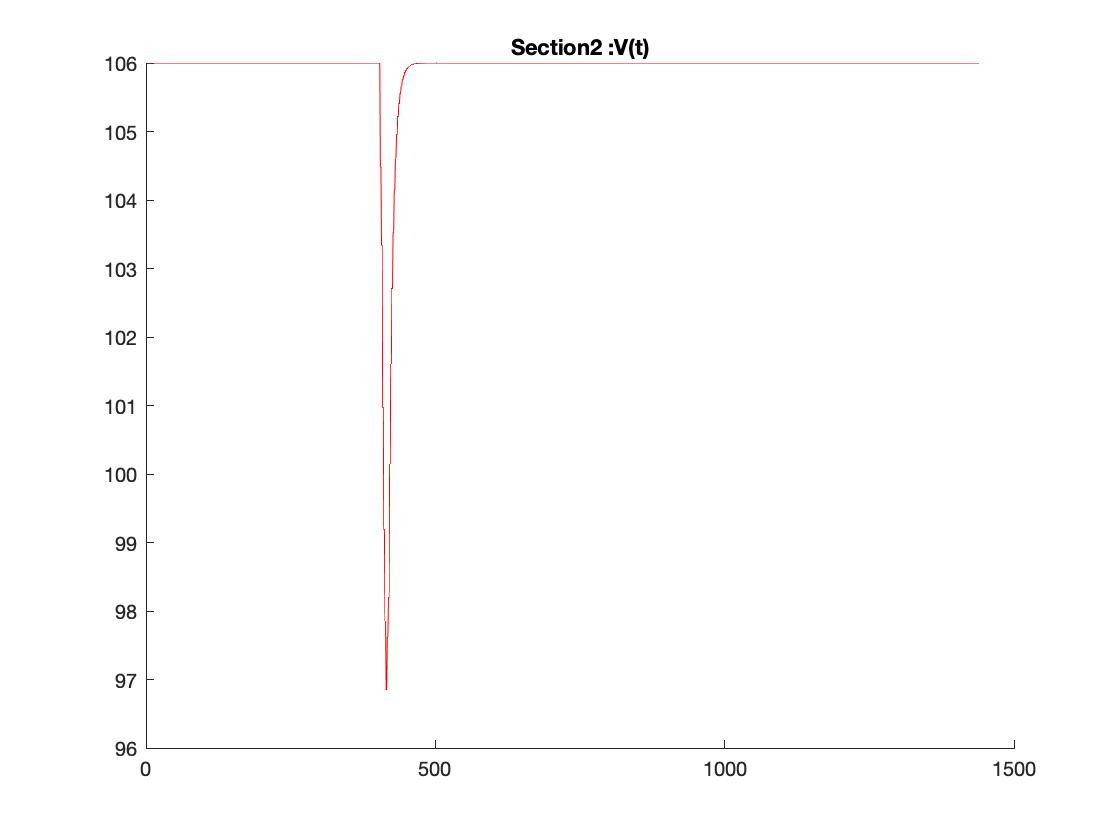,width=0.3\textwidth}}}
\caption{Section 2}\label{CSM2}
\end{figure*}

\begin{figure*}[!h]
\centering
\subfigure[Flow in \textit{[Veh/min]}]{\rotatebox{-0}{\epsfig{figure=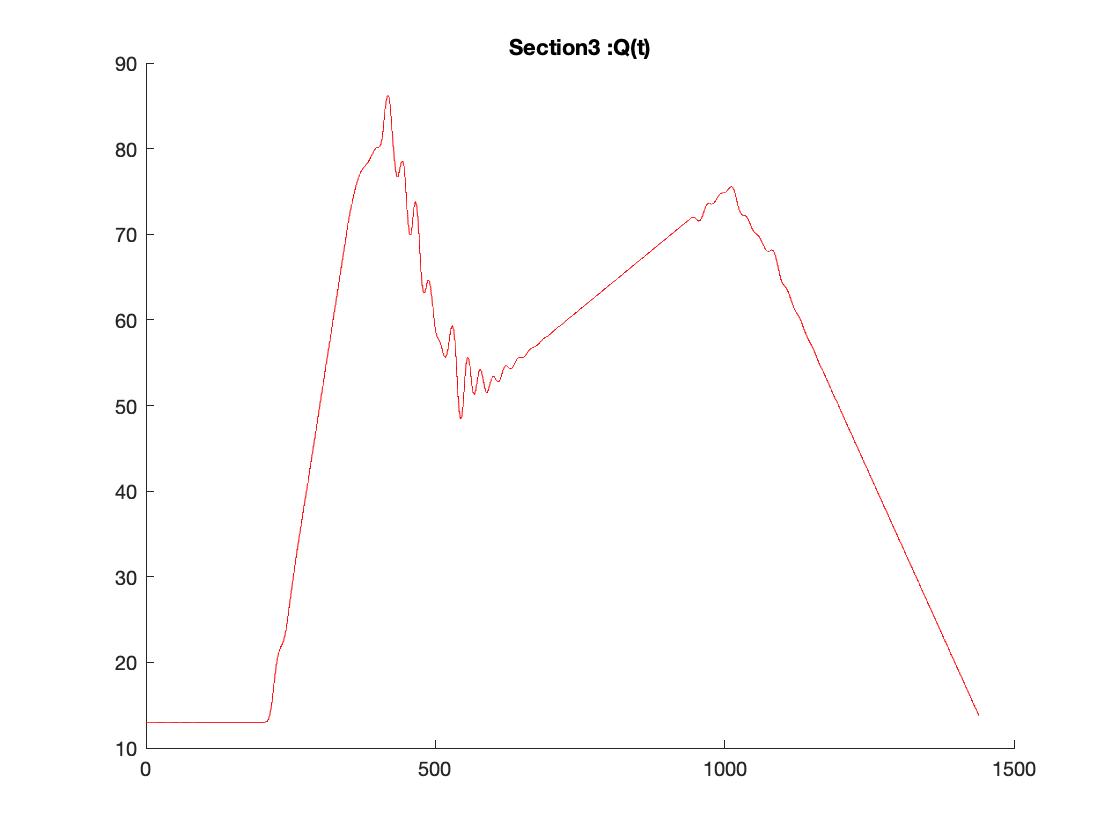,width=0.3\textwidth}}}
\subfigure[Occupancy rate in \textit{\%}]{\rotatebox{-0}{\epsfig{figure=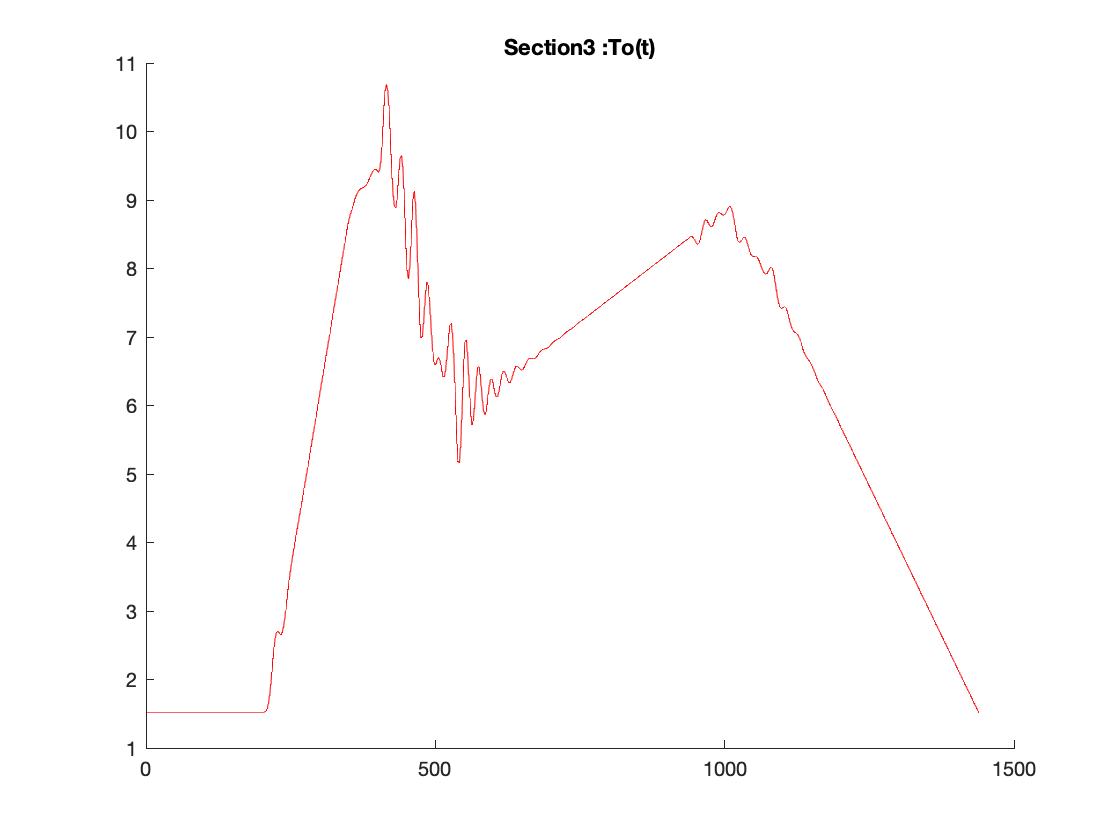,width=0.3\textwidth}}}
\subfigure[Speed in \textit{[km/h]}]{\rotatebox{-0}{\epsfig{figure=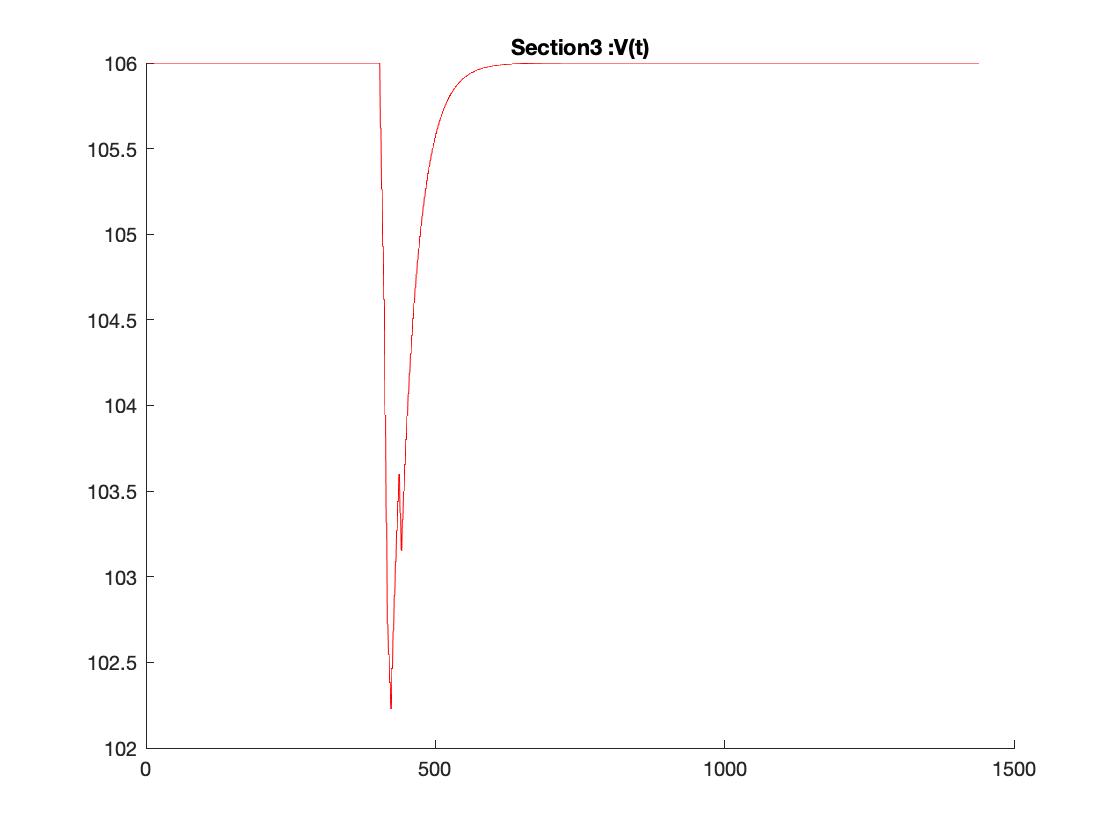,width=0.3\textwidth}}}
\caption{Section 3}\label{CSM3}
\end{figure*}

\begin{figure*}[!h]
\centering
\subfigure[Flow in \textit{[Veh/min]}]{\rotatebox{-0}{\epsfig{figure=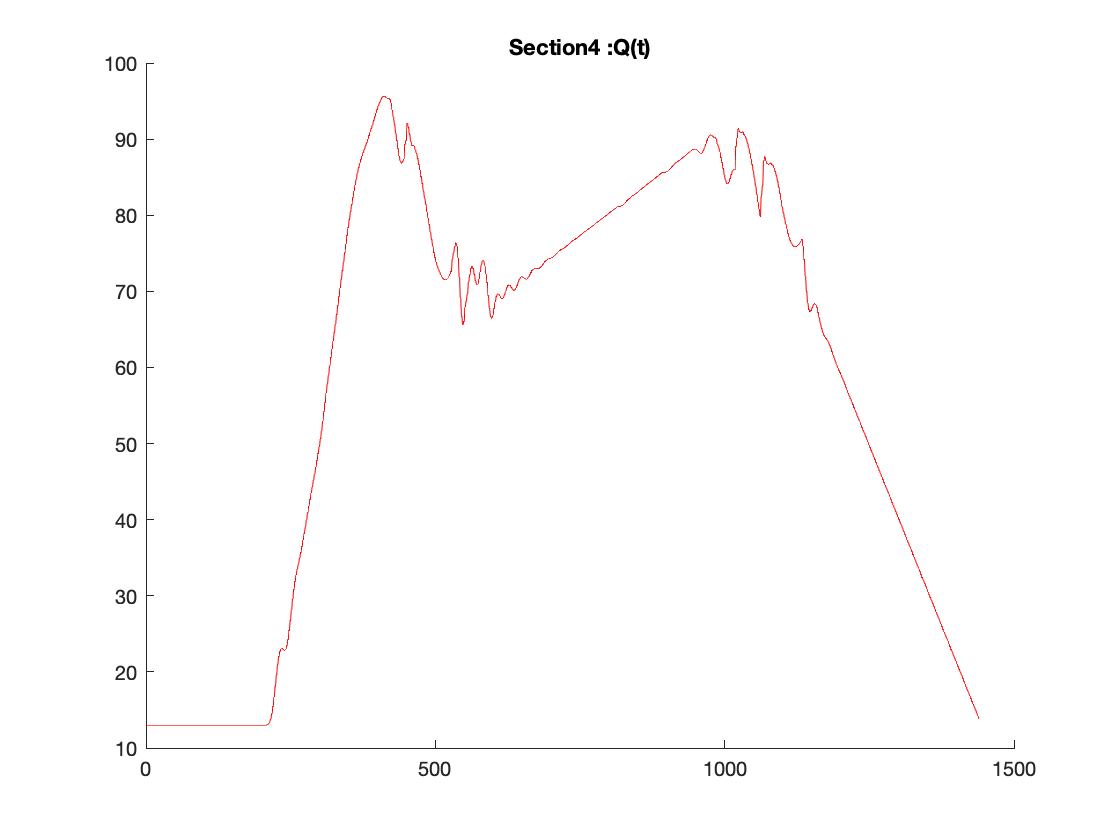,width=0.3\textwidth}}}
\subfigure[Occupancy rate in \textit{\%}]{\rotatebox{-0}{\epsfig{figure=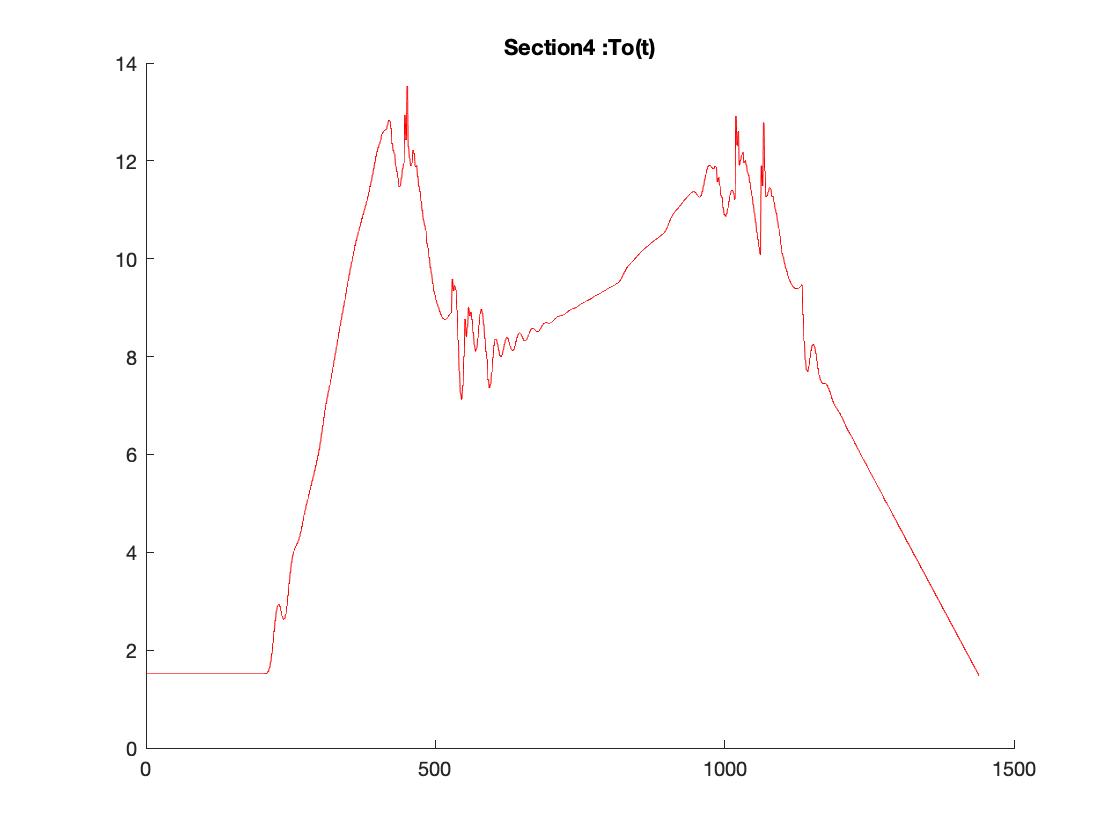,width=0.3\textwidth}}}
\subfigure[Speed in \textit{[km/h]}]{\rotatebox{-0}{\epsfig{figure=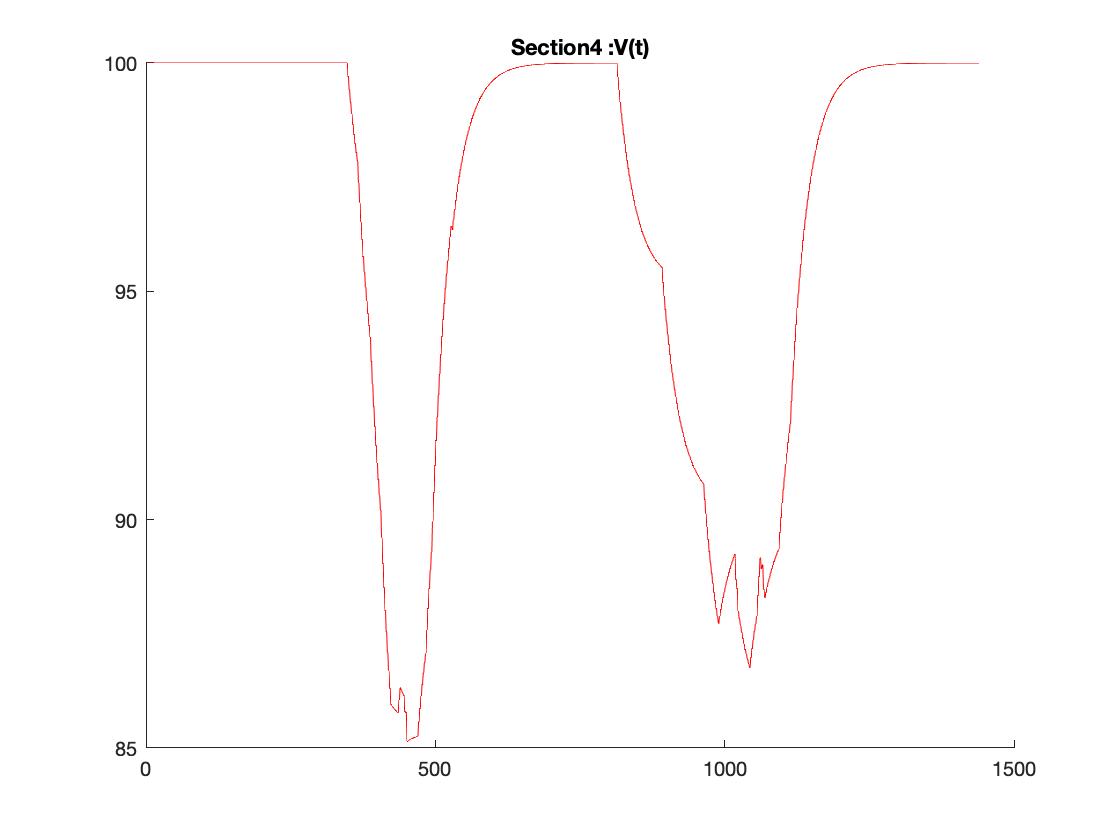,width=0.3\textwidth}}}
\caption{Section 4}\label{CSM4}
\end{figure*}

\begin{figure*}[!h]
\centering
\subfigure[Flow in \textit{[Veh/min]}]{\rotatebox{-0}{\epsfig{figure=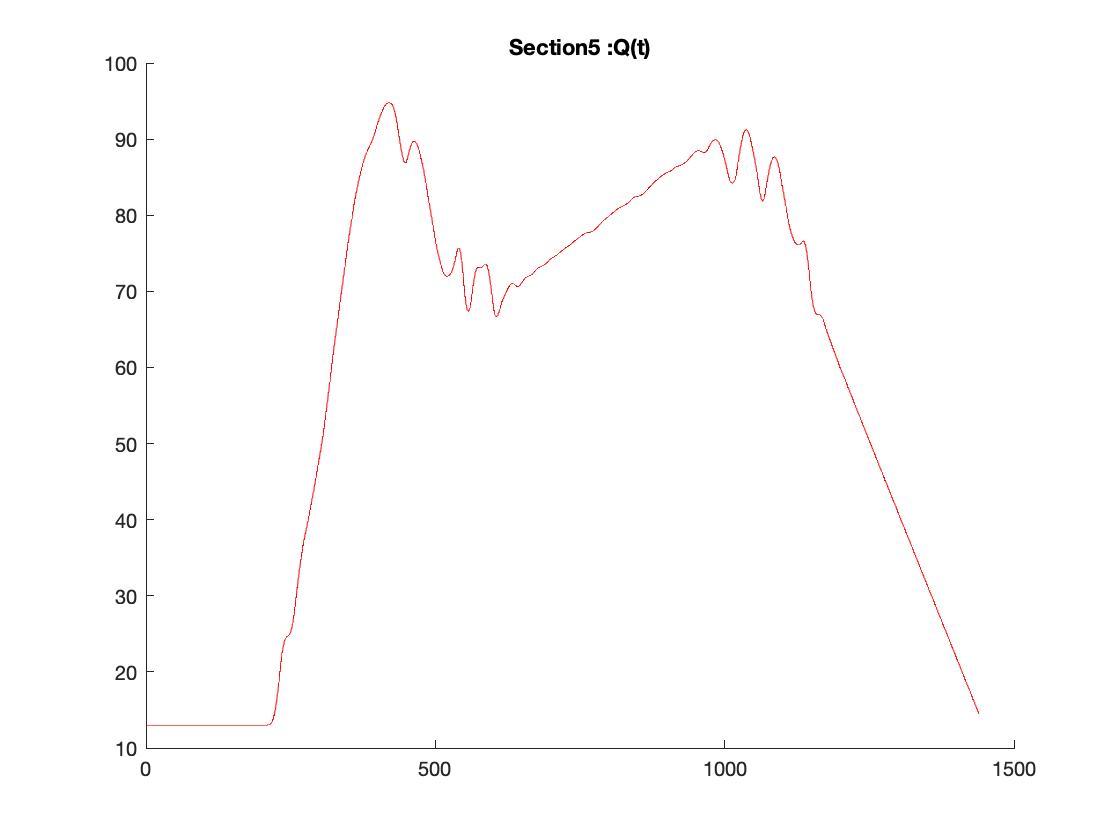,width=0.3\textwidth}}}
\subfigure[Occupancy rate in \textit{\%}]{\rotatebox{-0}{\epsfig{figure=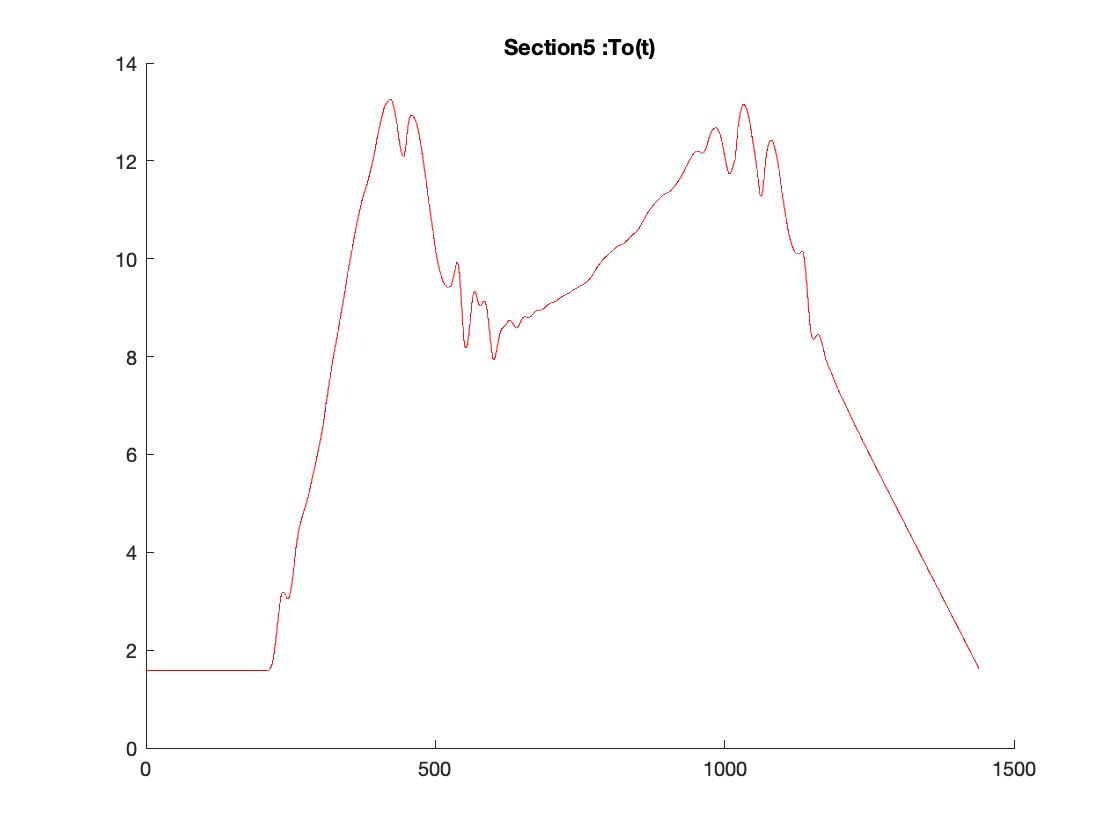,width=0.3\textwidth}}}
\subfigure[Speed in \textit{[km/h]}]{\rotatebox{-0}{\epsfig{figure=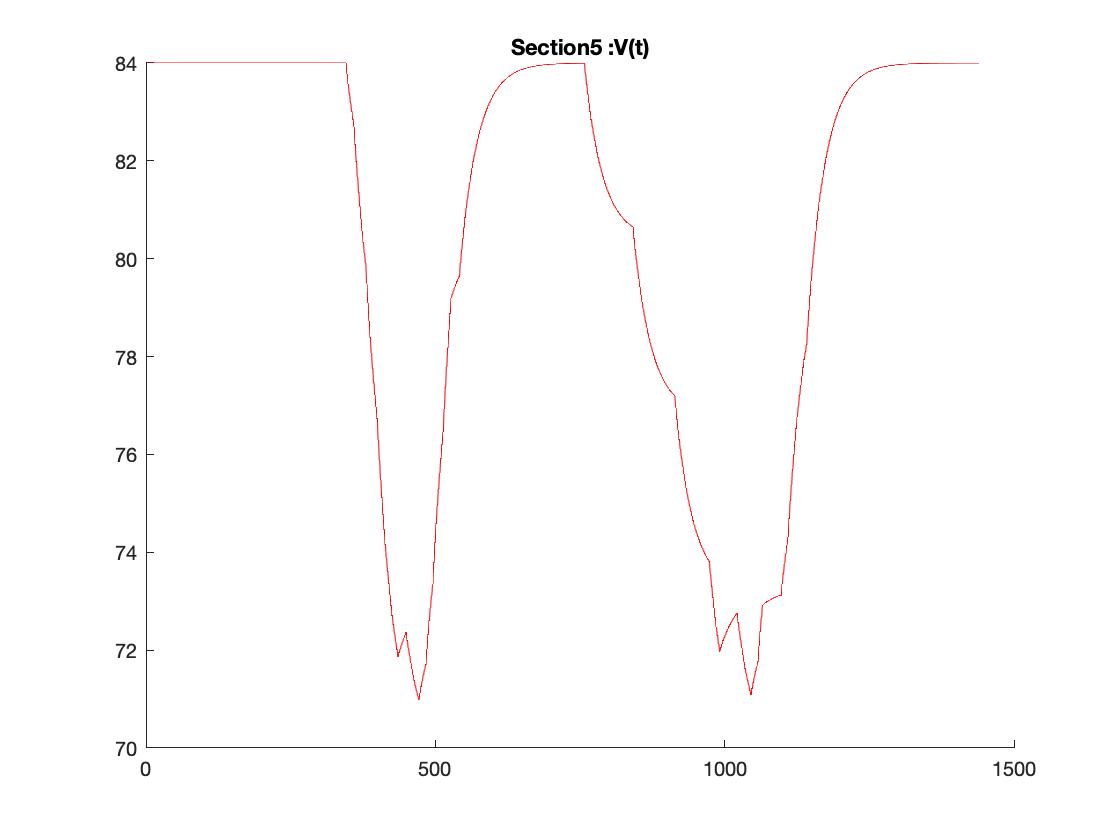,width=0.3\textwidth}}}
\caption{Section 5}\label{CSM5}
\end{figure*}

\begin{figure*}[!h]
\centering
\subfigure[Flow in \textit{[Veh/min]}]{\rotatebox{-0}{\epsfig{figure=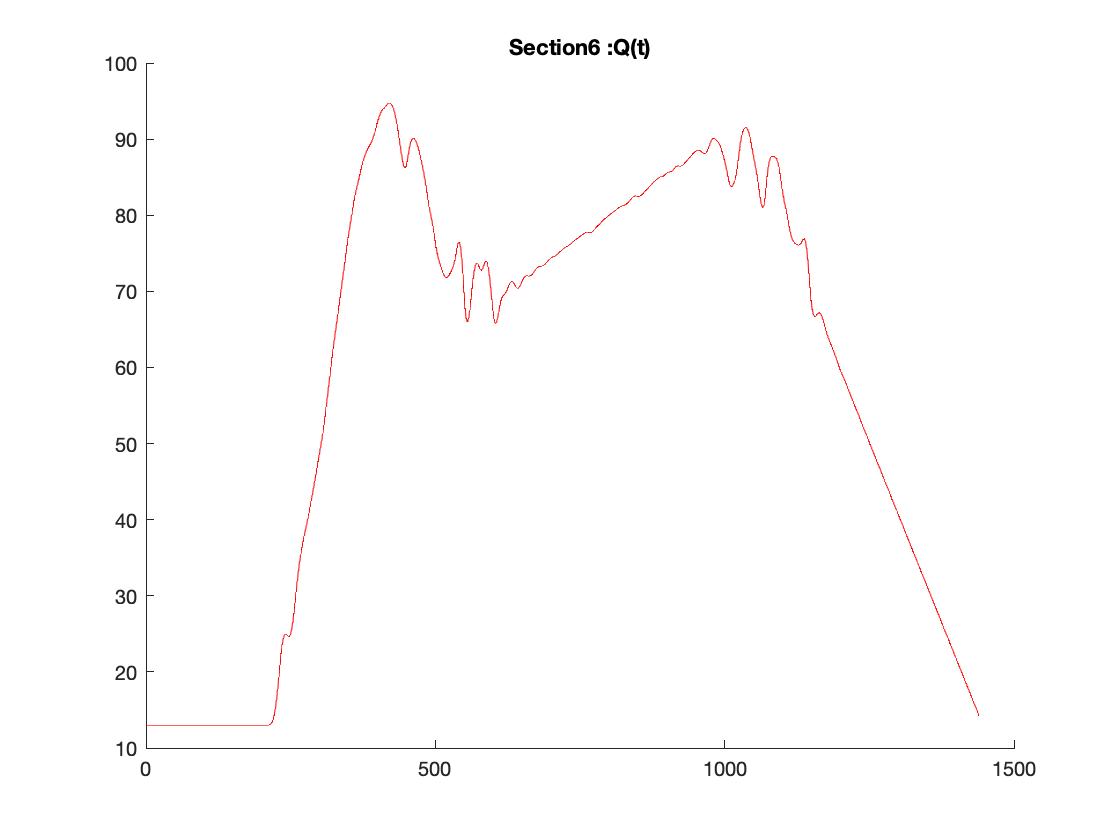,width=0.3\textwidth}}}
\subfigure[Occupancy rate in \textit{\%}]{\rotatebox{-0}{\epsfig{figure=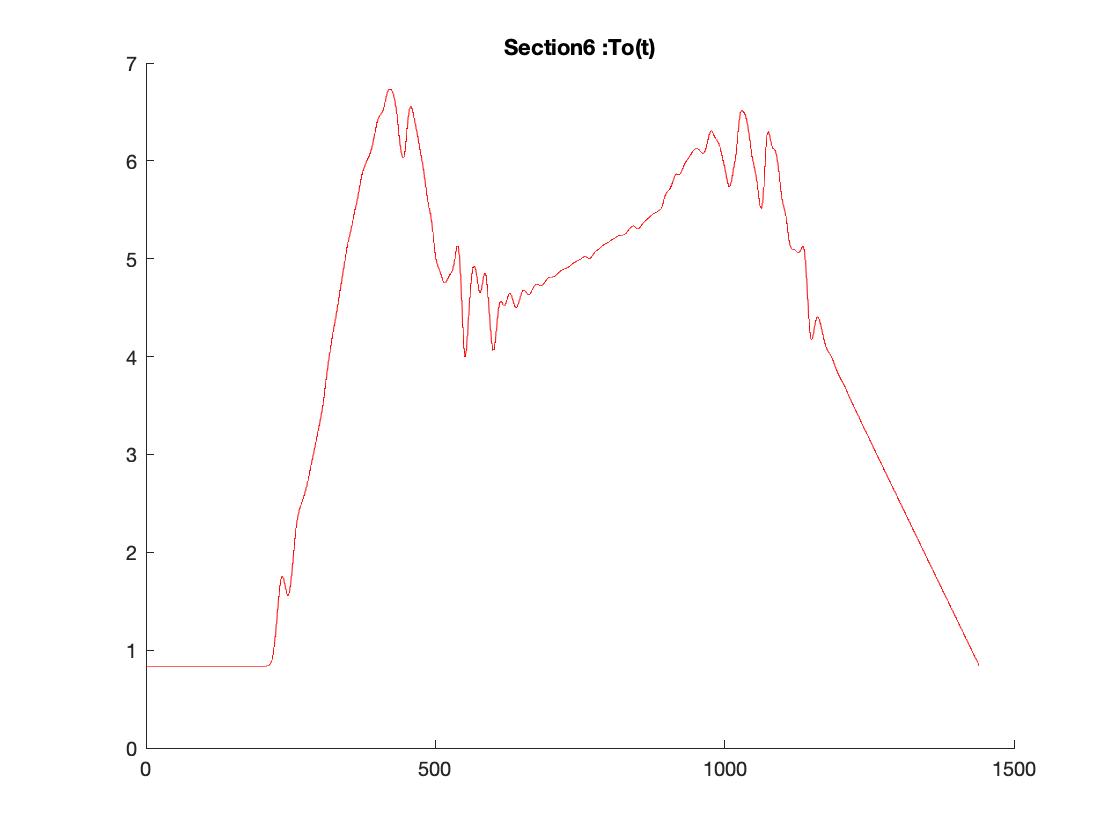,width=0.3\textwidth}}}
\subfigure[Speed in \textit{[km/h]}]{\rotatebox{-0}{\epsfig{figure=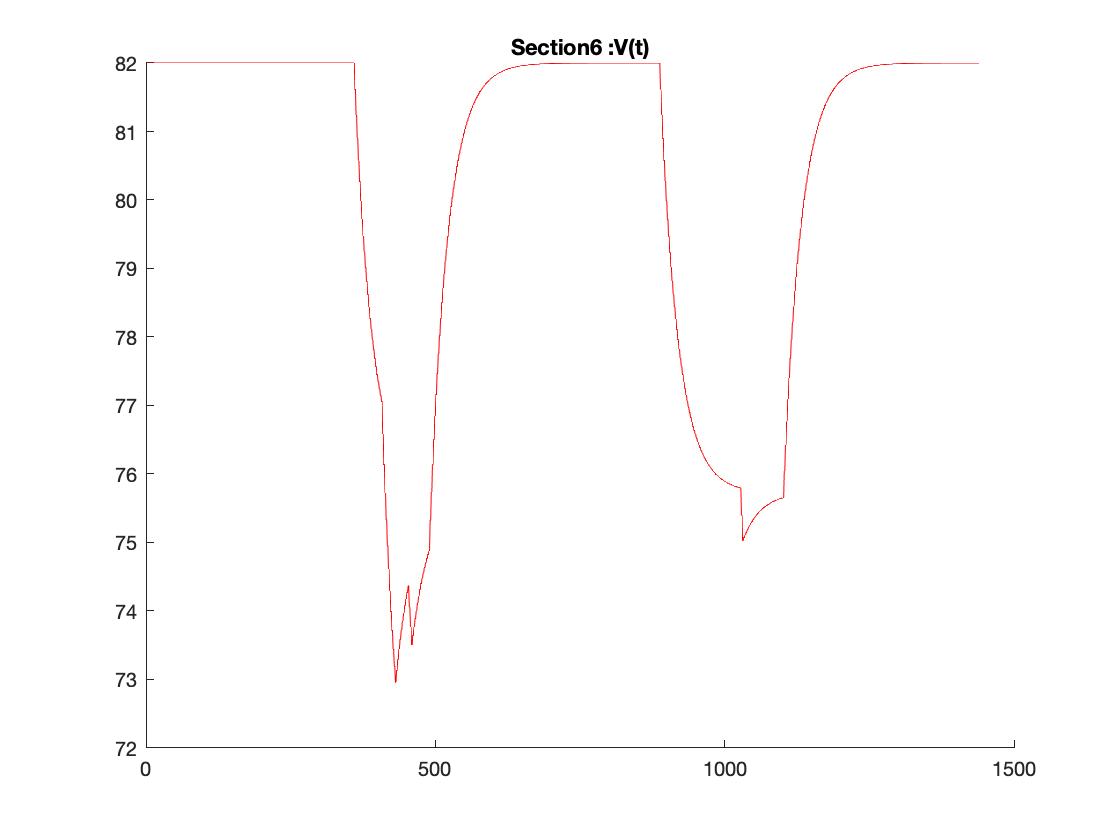,width=0.3\textwidth}}}
\caption{Section 6}\label{CSM6}
\end{figure*}

\begin{figure*}[!h]
\centering
\subfigure[Flow in \textit{[Veh/min]}]{\rotatebox{-0}{\epsfig{figure=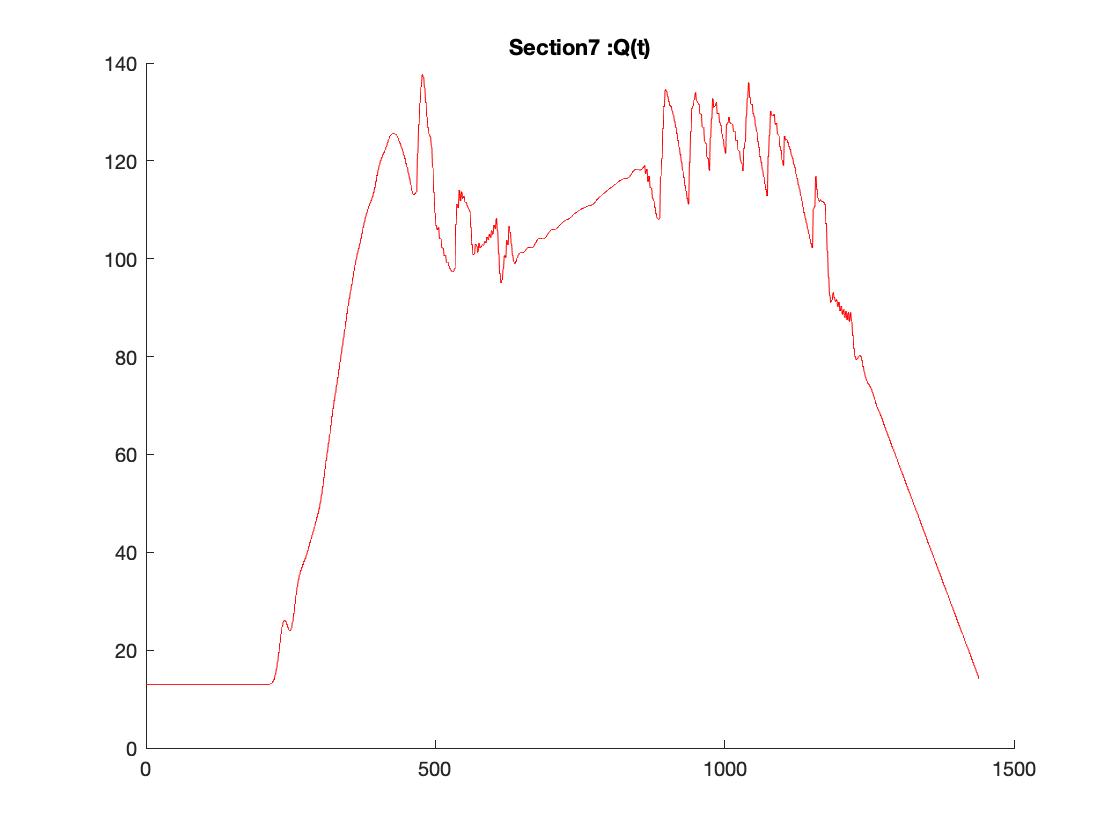,width=0.3\textwidth}}}
\subfigure[Occupancy rate in \textit{\%}]{\rotatebox{-0}{\epsfig{figure=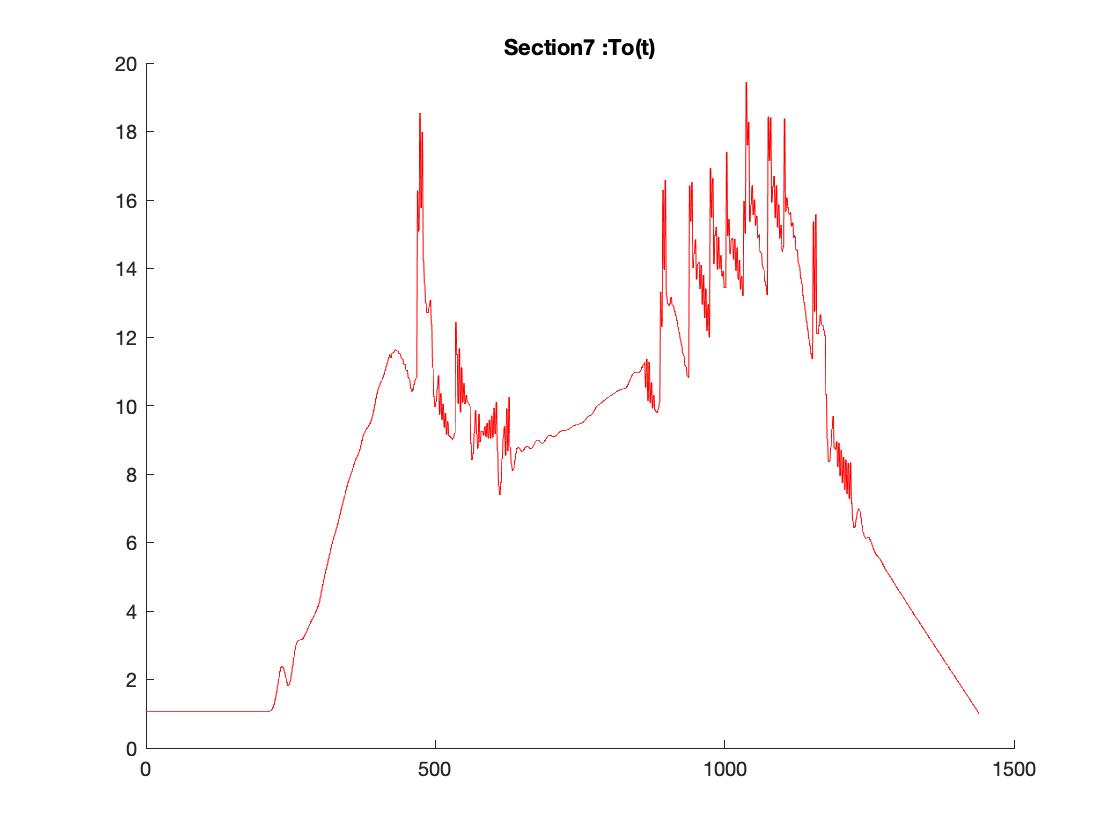,width=0.3\textwidth}}}
\subfigure[Speed in \textit{[km/h]}]{\rotatebox{-0}{\epsfig{figure=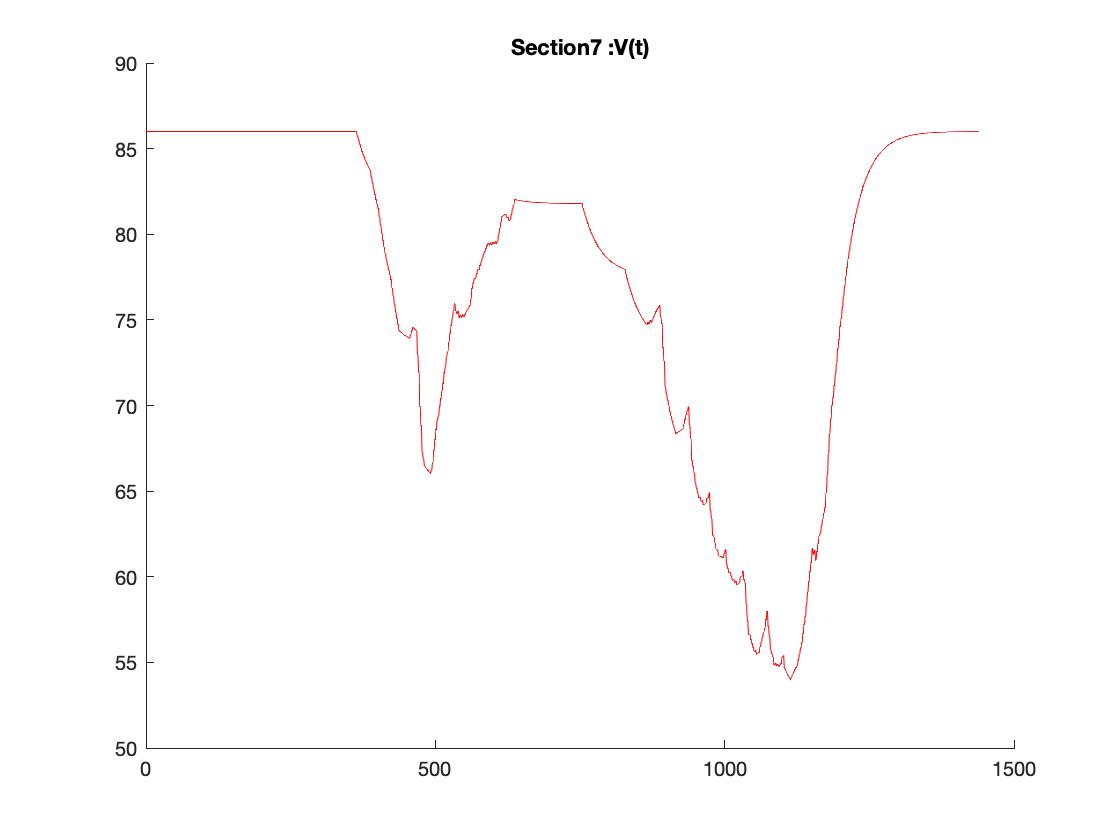,width=0.3\textwidth}}}
\caption{Section 7}\label{CSM7}
\end{figure*}

\begin{figure*}[!h]
\centering
\subfigure[Ramp 1]{\rotatebox{-0}{\epsfig{figure=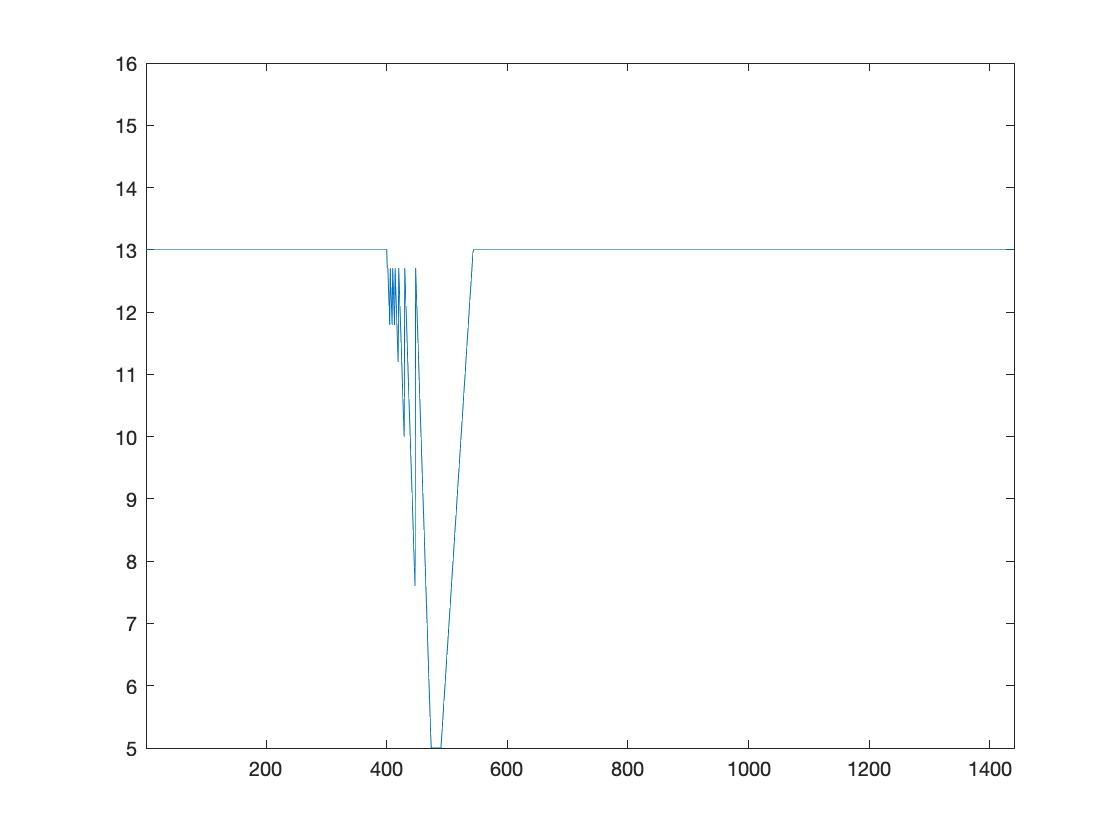,width=0.3\textwidth}}}
\subfigure[Ramp 2]{\rotatebox{-0}{\epsfig{figure=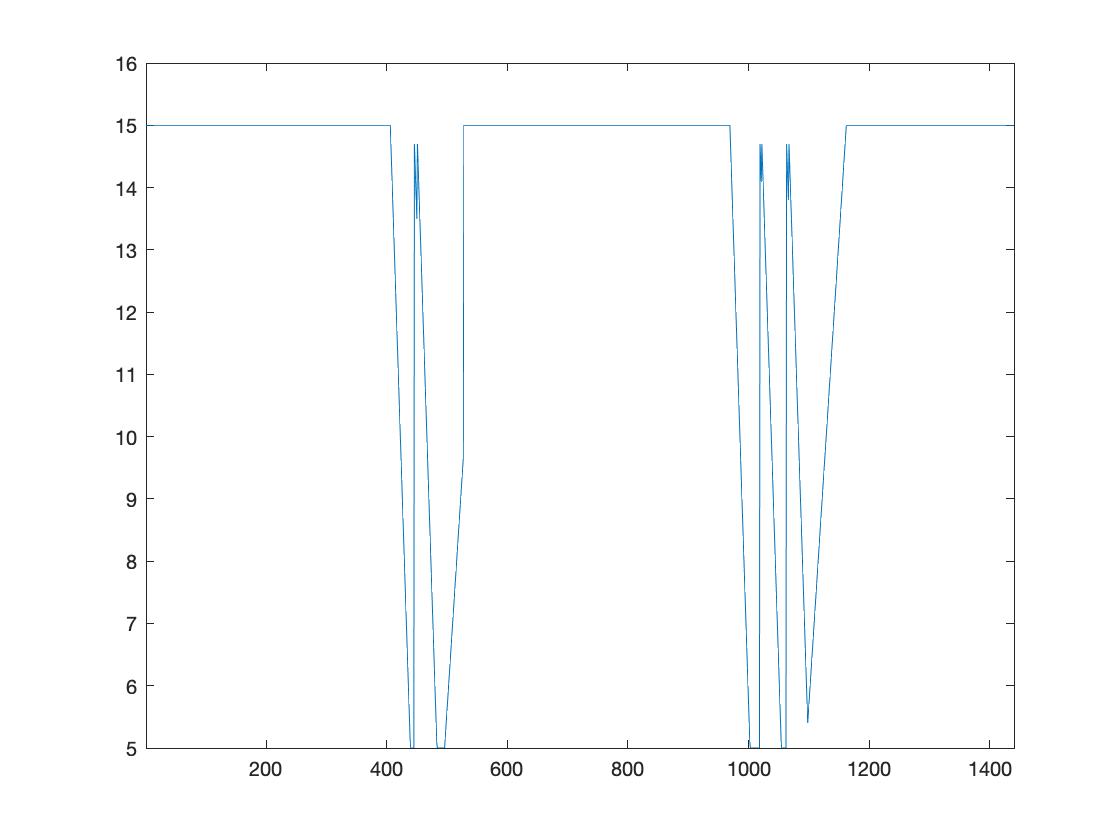,width=0.3\textwidth}}}
\subfigure[Ramp 3]{\rotatebox{-0}{\epsfig{figure=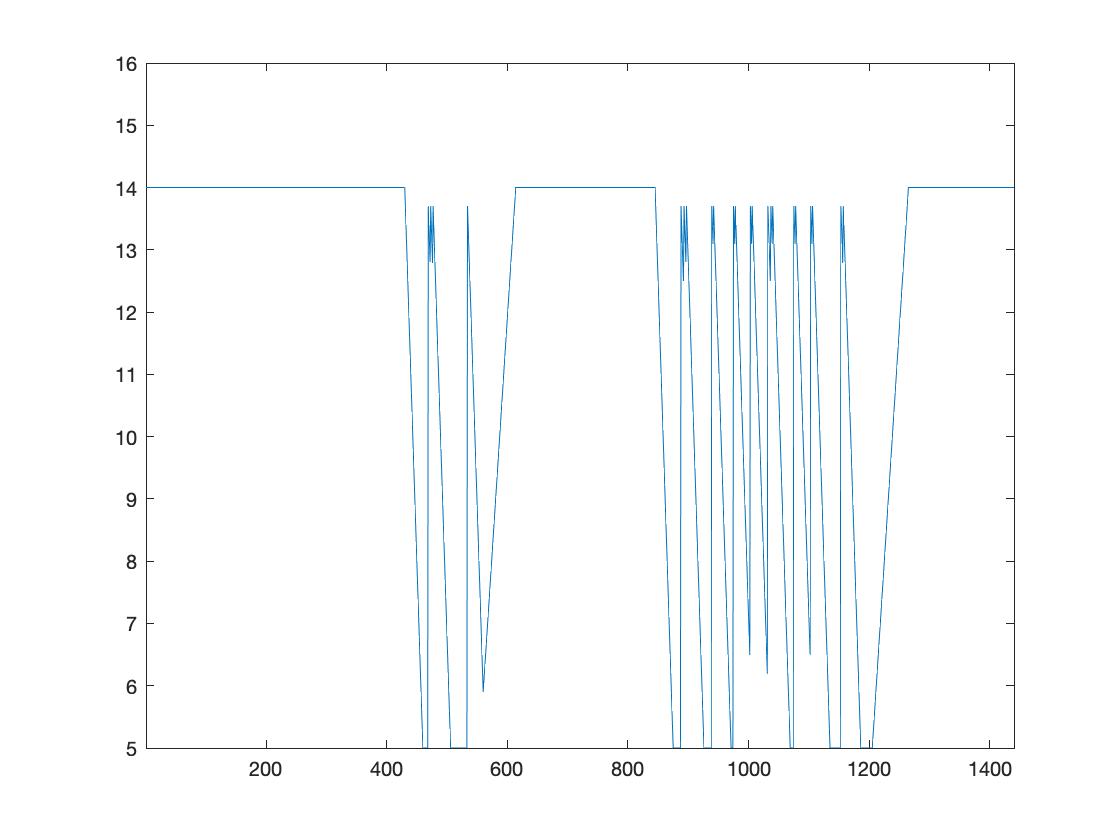,width=0.3\textwidth}}}
\caption{Setpoint time evolution}\label{C}
\end{figure*}

\begin{figure*}[!h]
\centering
\subfigure[Ramp 1]{\rotatebox{-0}{\epsfig{figure=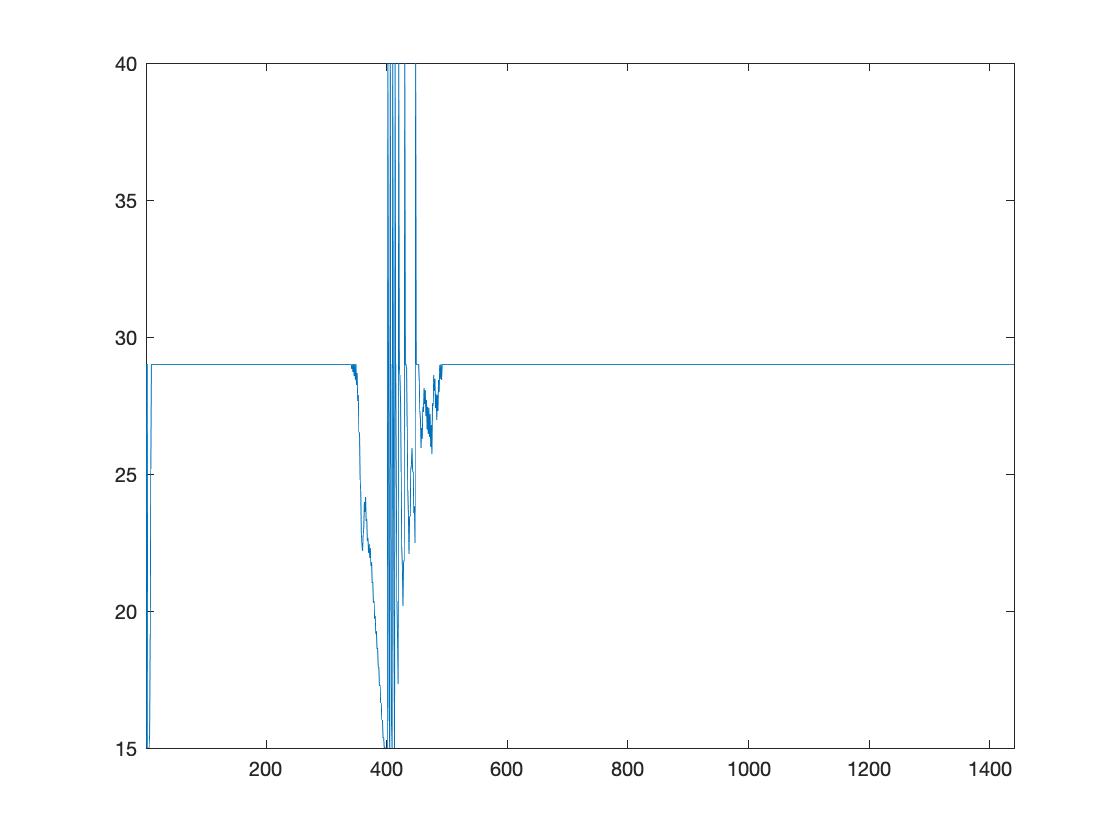,width=0.3\textwidth}}}
\subfigure[Ramp 2]{\rotatebox{-0}{\epsfig{figure=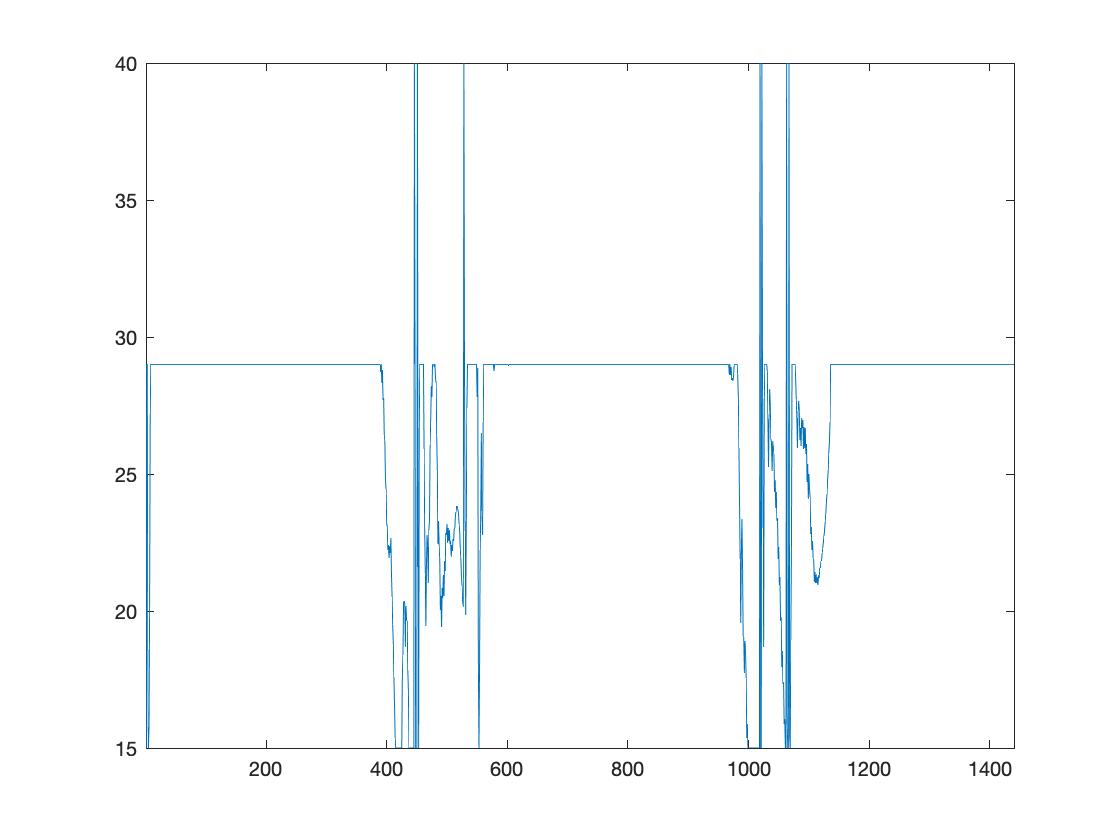,width=0.3\textwidth}}}
\subfigure[Ramp 3]{\rotatebox{-0}{\epsfig{figure=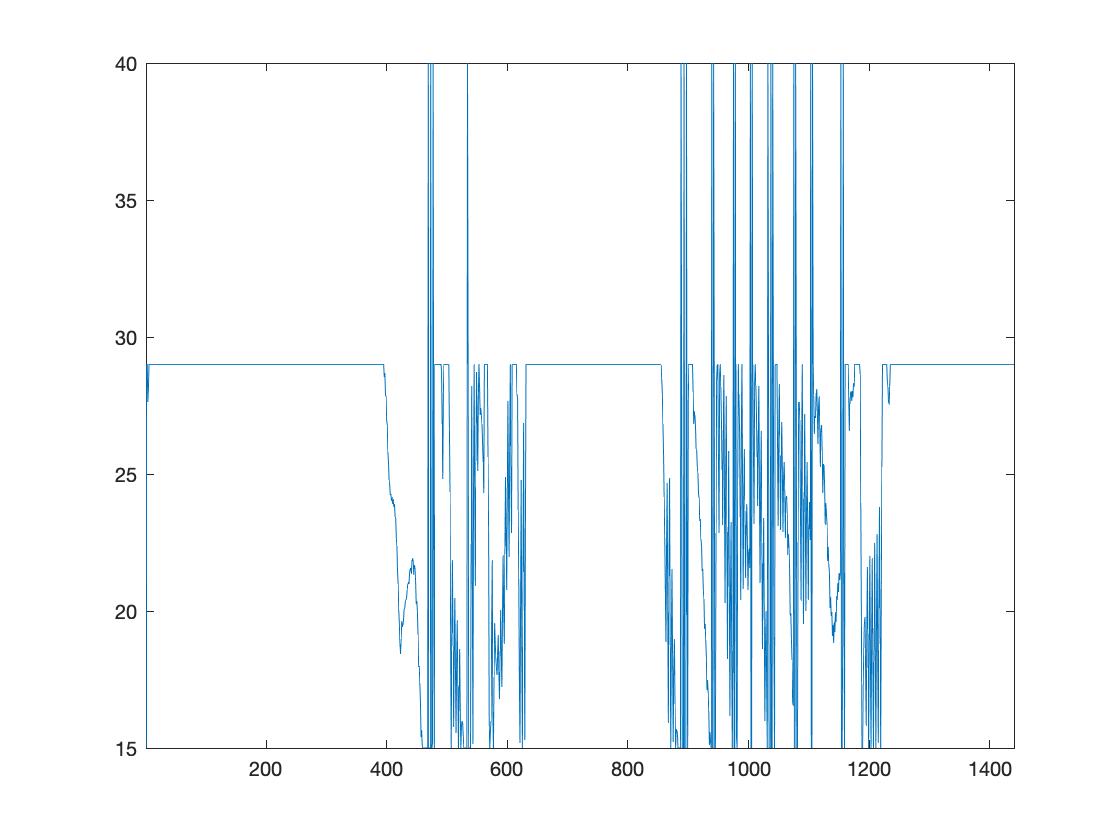,width=0.3\textwidth}}}
\caption{Time evolution of green lights according to the control inputs $GD_1$, $GD_2$, $GD_3$ }\label{V}
\end{figure*}

\begin{figure*}[!h]
\centering
\subfigure[Ramp 1]{\rotatebox{-0}{\epsfig{figure=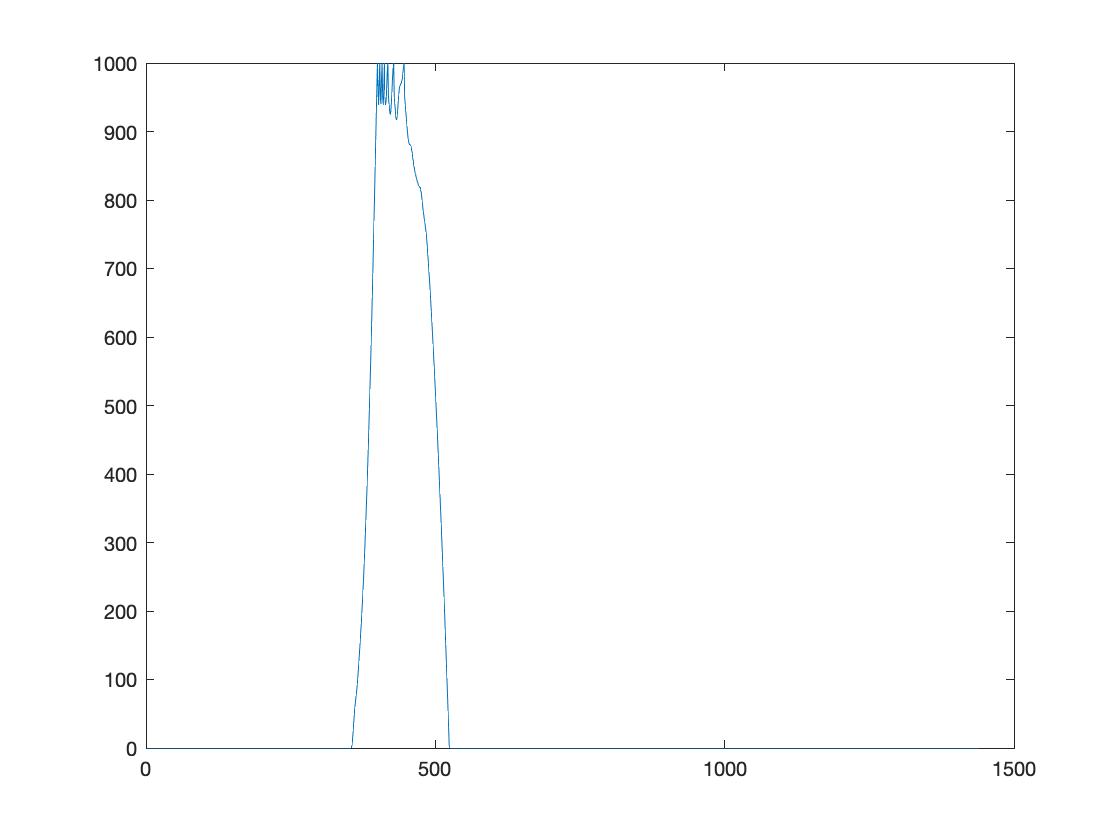,width=0.3\textwidth}}}
\subfigure[Ramp 2]{\rotatebox{-0}{\epsfig{figure=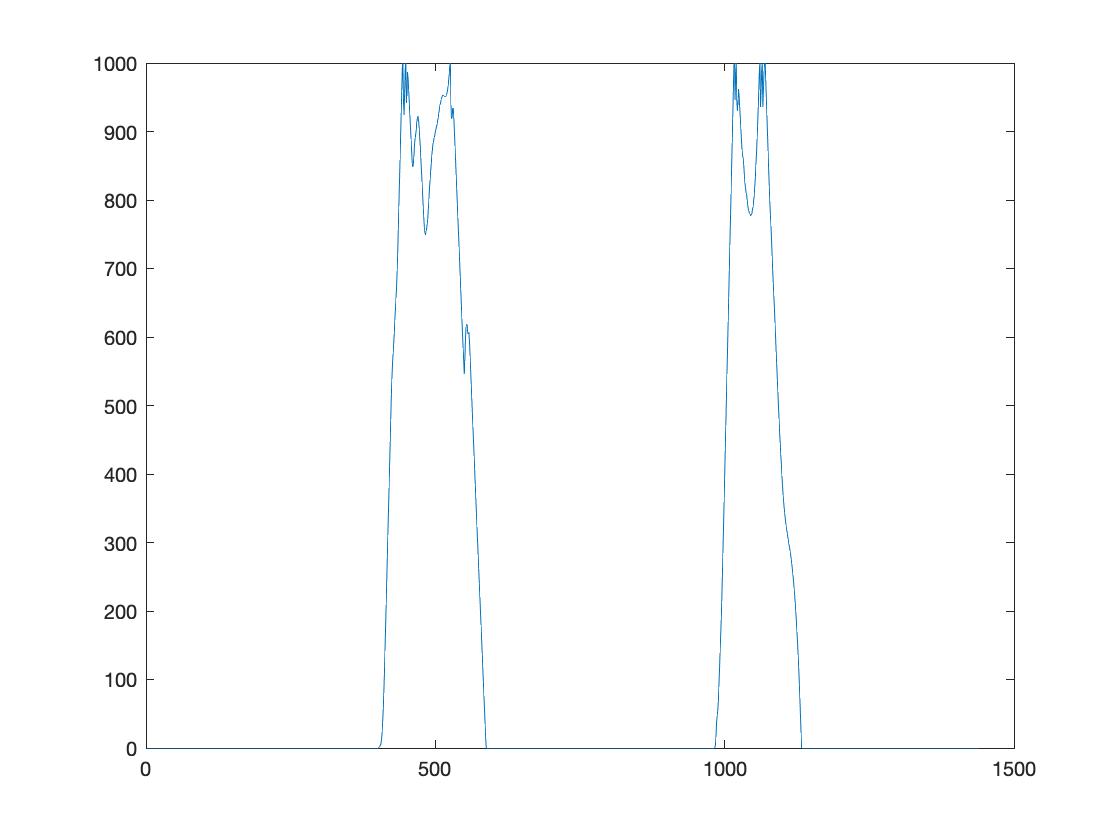,width=0.3\textwidth}}}
\subfigure[Ramp 3]{\rotatebox{-0}{\epsfig{figure=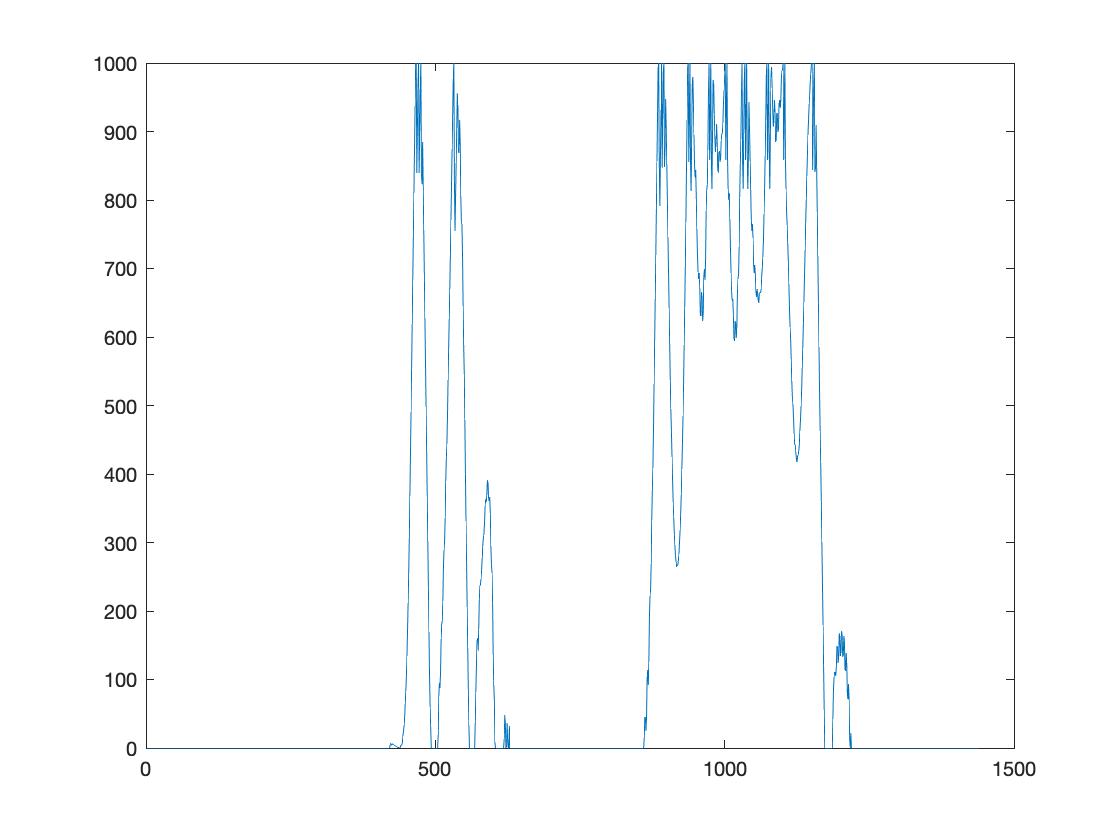,width=0.3\textwidth}}}
\caption{Queue length in [m]}\label{L}
\end{figure*}

\begin{figure*}[!h]
\centering
\subfigure[Ramp 1]{\rotatebox{-0}{\epsfig{figure=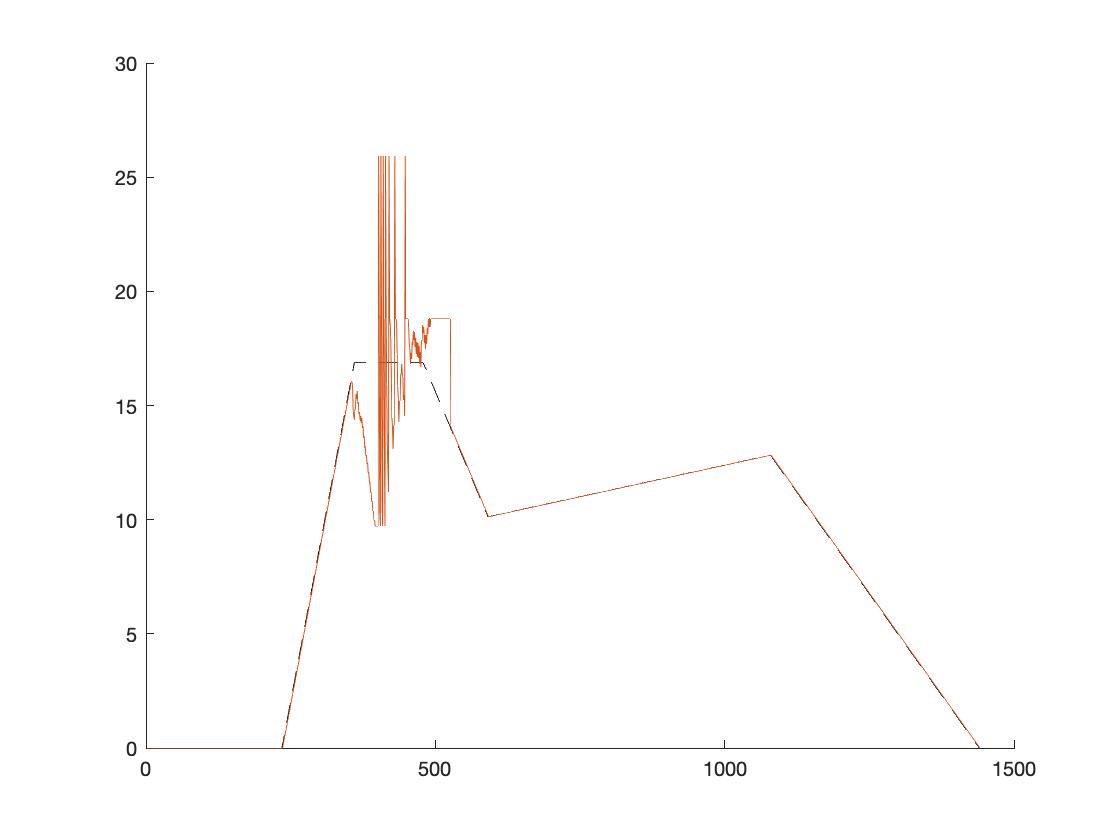,width=0.3\textwidth}}}
\subfigure[Ramp 2]{\rotatebox{-0}{\epsfig{figure=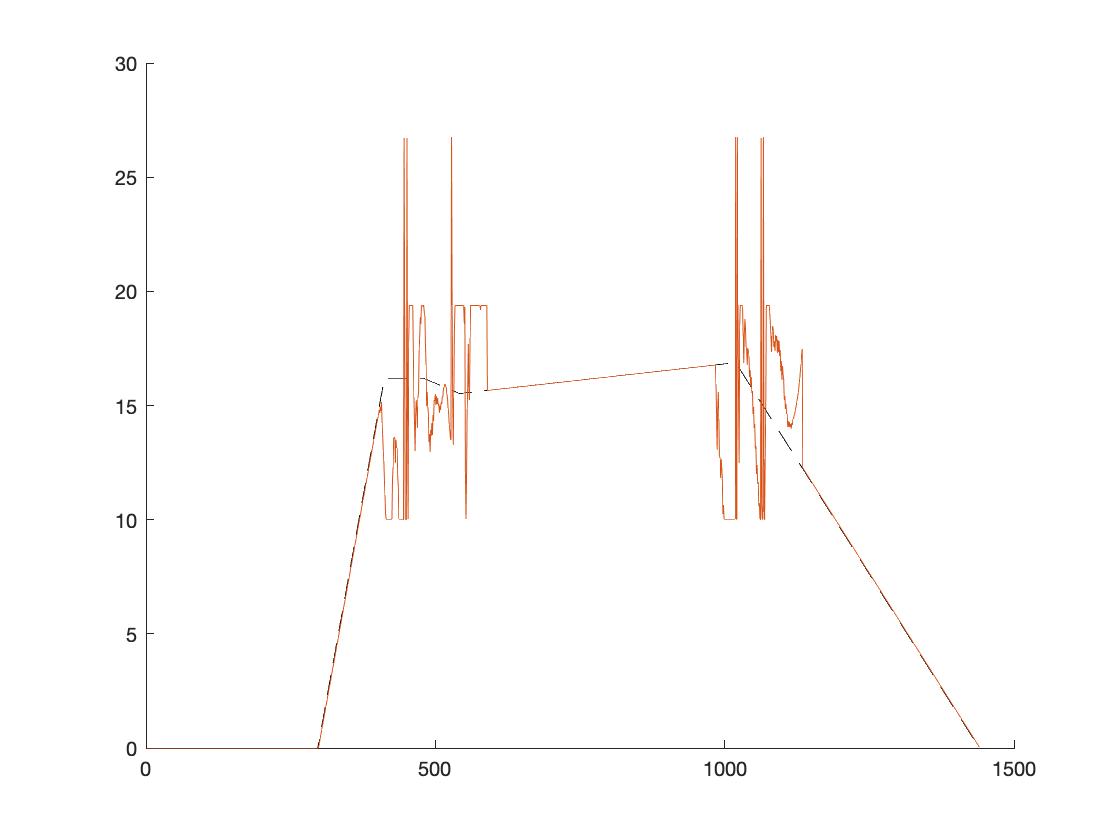,width=0.3\textwidth}}}
\subfigure[Ramp 3]{\rotatebox{-0}{\epsfig{figure=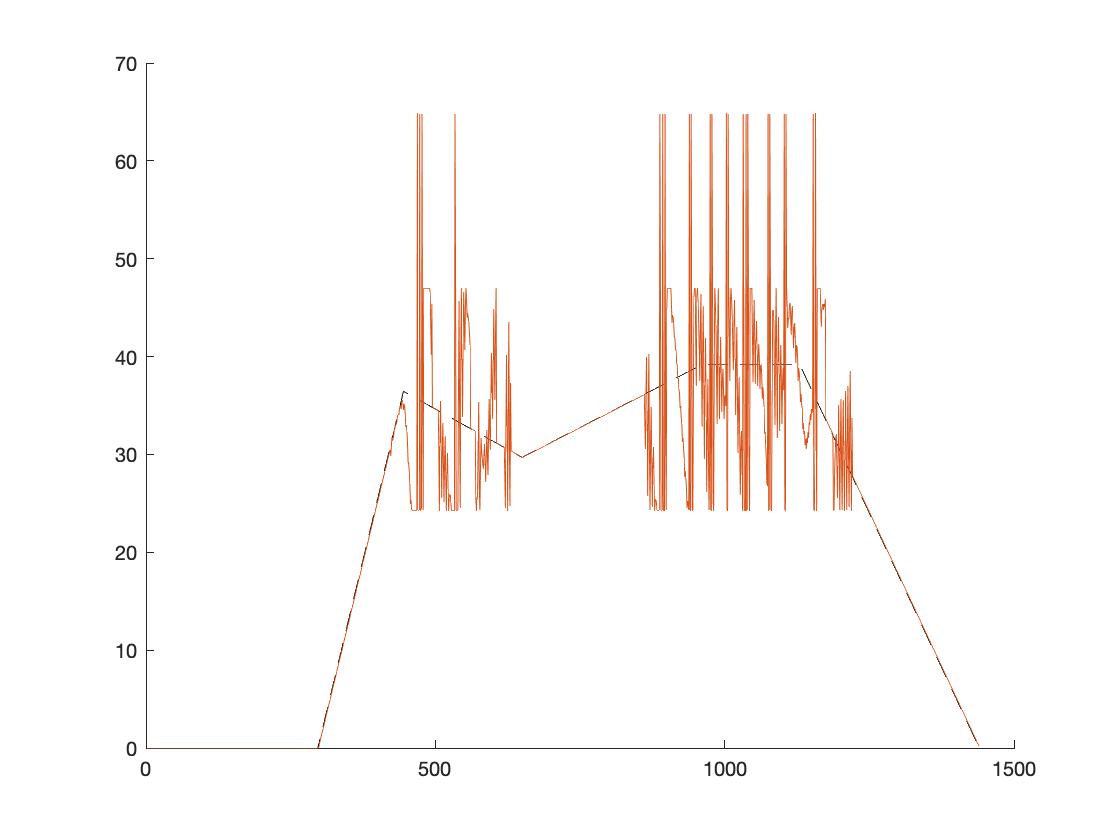,width=0.3\textwidth}}}
\caption{Demand $D_1,j$ (- -) and flow $Q_1,j$ (--)  in \textit{[Veh/min]}}\label{Q}
\end{figure*}

Figures  \ref{CSM1} - \ref{Q} display convincing results via the setting of Section \ref{iptraffic}:
\begin{itemize}
\item The reference trajectory is a decreasing time function when the traffic is dense (see Figure \ref{C}). This property does not hold anymore if the traffic becomes fluid.
\item Figure \ref{V} shows that the green light duration is set to the whole cycle of 40s when the queue on the ramp is large (see Figure \ref{L}). 
\item Figure \ref{Q} indicates that the output flow on the ramps are reduced when there is a congestion. 

\end{itemize}
\begin{remark}
Those results are only slightly better than those obtained via ALINEA with the same calibration of the setpoint. Let us emphasize however that ALINEA when implemented with the critical density $\rho_c$ gives results which are disapointly similar to those obtained without any control, \textit{i.e.}, without ramp metering.
\end{remark}


\subsection{What is happening with a sudden change?}\label{sudd}
\begin{figure*}[!h]
\centering
\subfigure[TTS]{\rotatebox{-0}{\epsfig{figure=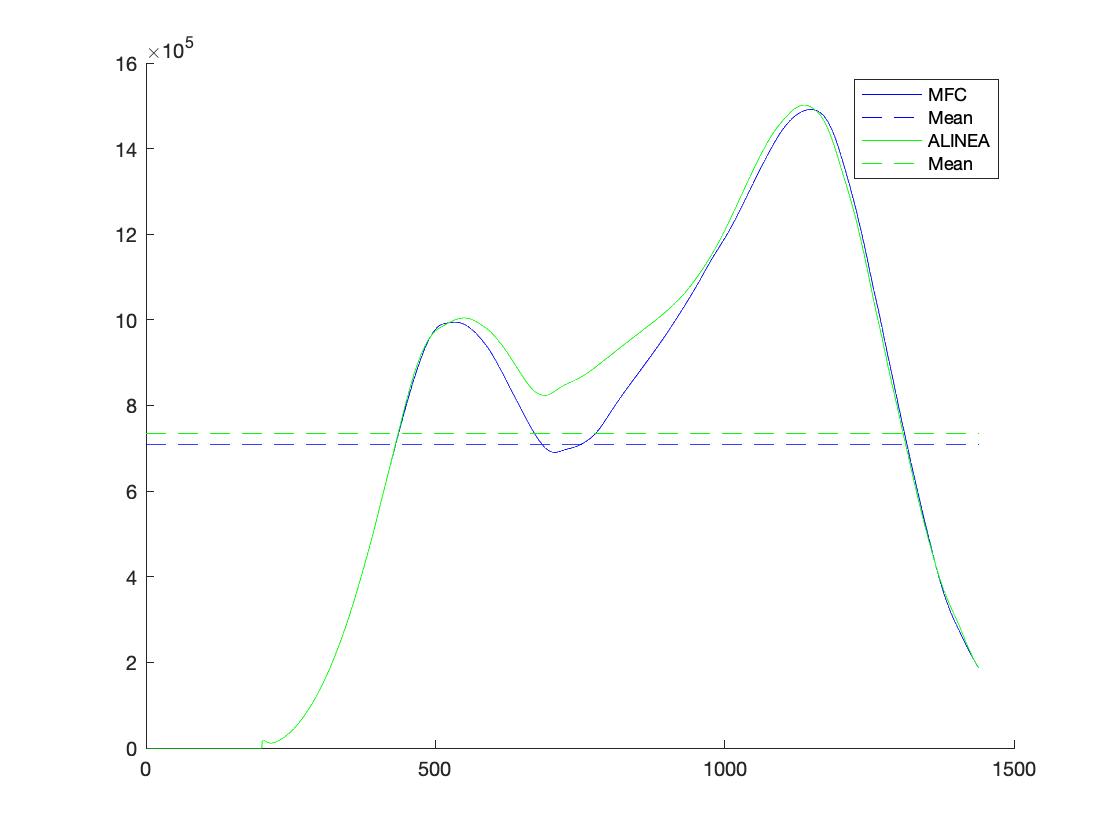,width=0.3\textwidth}}}
\subfigure[TTP]{\rotatebox{-0}{\epsfig{figure=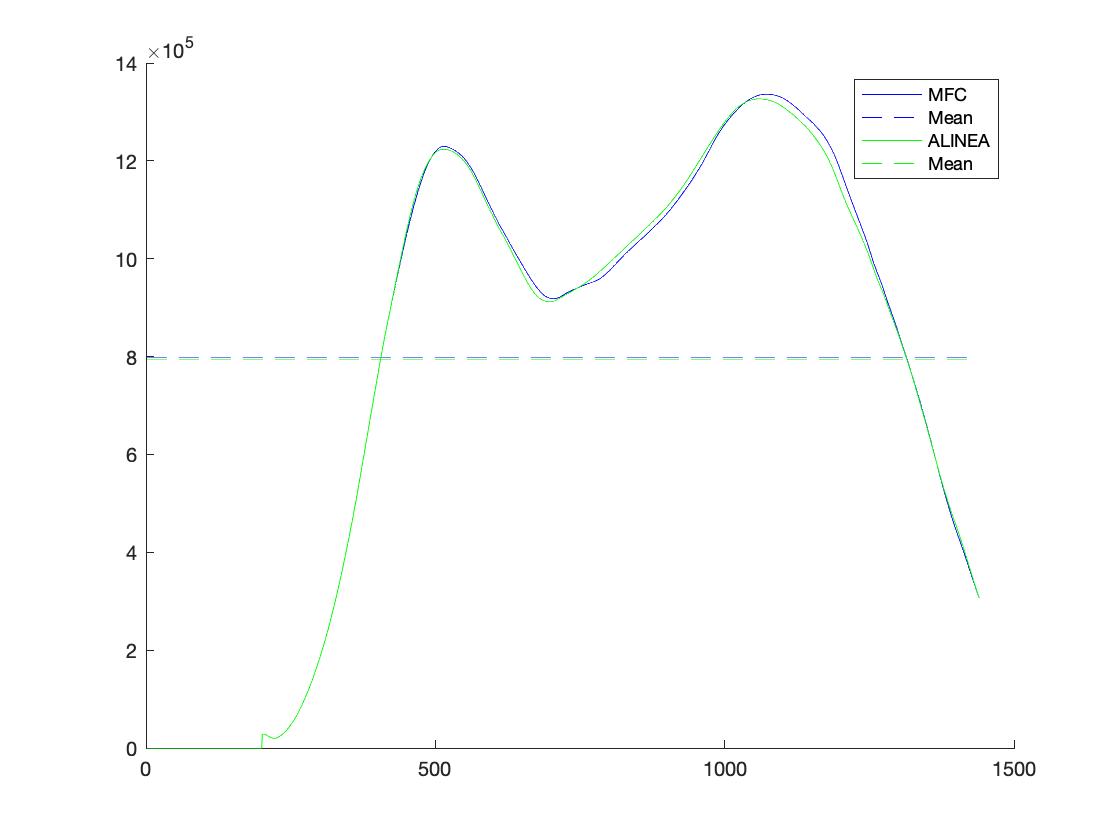,width=0.3\textwidth}}}
\subfigure[MS]{\rotatebox{-0}{\epsfig{figure=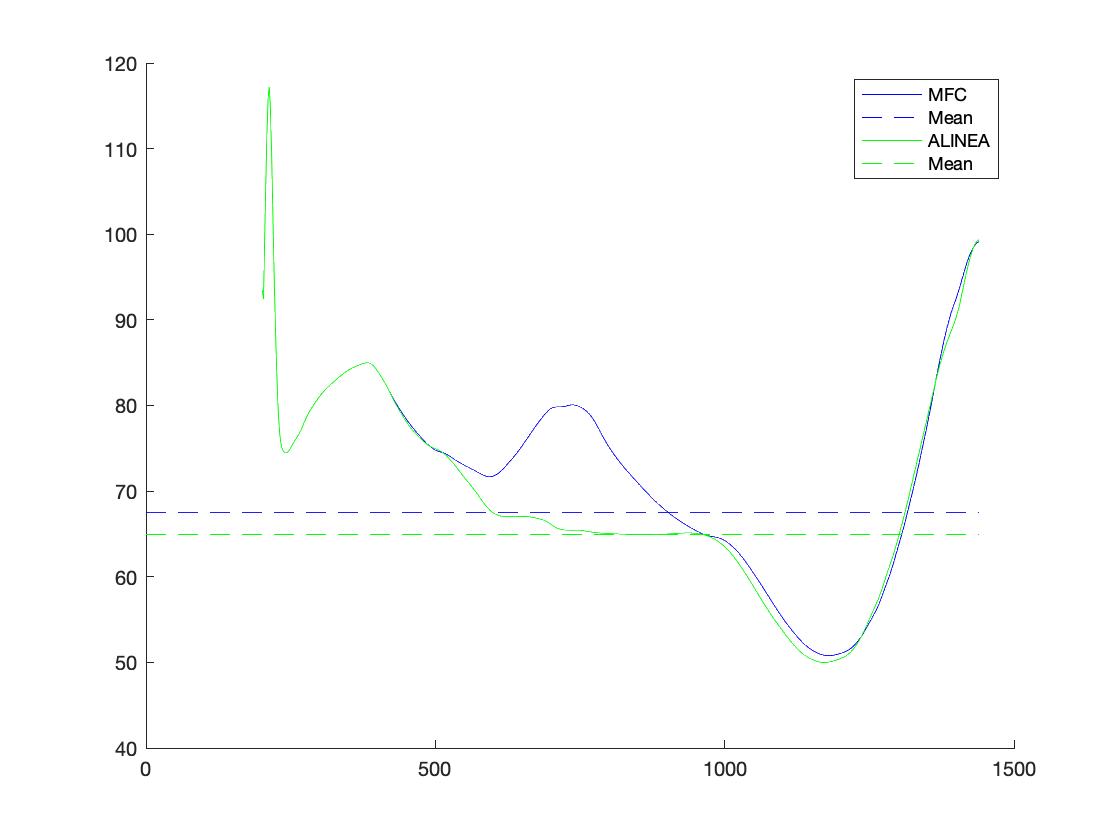,width=0.3\textwidth}}}
\caption{ALINEA and MFC \newline (TTS: Travel Time Spent, TTD: Total Travel Distance, MS: Mean Speed) }\label{P4}
\end{figure*}

Assume only $2$ lanes, instead of $3$, on Segment $S_2$: it might be due to an accident or to some work on the highway. Figures \ref{P4} shows that the iP, without any new tuning, behave much better than ALINEA, especially outside the congestion hours when ramp metering is more or less useless.  Those results agree with Section \ref{prop}. 

\section{Conclusion}\label{conclu}
Our approach has already been successfully employed on several French highways. Its adaptative features explains its excellent robustness with respect to unexpected events. Its implementation moreover is quite elementary.
There are of course other ramp metering settings which ought to be compared with our viewpoint, especially now those stemming from artificial intelligence. Le us notice that \emph{deep reinforcement learning} methods do not seem today to perform better than ALINEA (see, \textit{e.g.}, \cite{bel}).

The \emph{coordination} of several ramp-metering actions on highway networks is  being investigated. Coordinated ALINEA techniques have been used in Australia some time ago \cite{coordin}. It should be clear that any such coordination ought to be connected to a \emph{variable speed limit} (see, \textit{e.g.}, \cite{khon}) in order to be more efficient.



\end{document}